\documentclass[article]{elsarticle}
\usepackage{lineno,hyperref}
\usepackage{subfigure}
\usepackage{placeins}
\usepackage{algorithm}
\usepackage{algorithmic}
\usepackage{amsmath}
\usepackage{amsmath,amsfonts,amssymb}
\usepackage{setspace}
\usepackage{xcolor}
\usepackage{hyperref}
\usepackage{cleveref}
\usepackage{natbib}
\let\Algorithm\algorithm
\renewcommand\algorithm[1][]{\Algorithm[#1]\setstretch{1.30}}
\newcommand{\quotes}[1]{``#1''}

\newcommand{\scapin}[1]{{\color{black}#1}}

\begin{document}
\begin{frontmatter}
%
%\title{TITLE: An interface capturing method for liquid-gas flows at low-Mach number}
\title{An interface capturing method for liquid-gas flows at low-Mach number}
%
%\author[unipd]{Federico Dalla Barba\corref{mycorrespondingauthor}}
\author[unipd]{Federico Dalla Barba}
\ead{federico.dallabarba@phd.unipd.it}
\author[kth]{Nicol\'o Scapin\corref{mycorrespondingauthor}}
\ead{nicolos@mech.kth.se}
\cortext[mycorrespondingauthor]{Corresponding author}
\author[oist]{Marco E. Rosti}
\ead{marco.rosti@oist.jp}
\author[unipd]{Francesco Picano}
\ead{francesco.picano@unipd.it}
\author[kth,ntnu]{Luca Brandt}
\ead{luca@mech.kth.se}
\address[unipd]{Department of Industrial Engineering \& CISAS, University of Padova,  Padova, Italy}
\address[oist]{Complex Fluids and Flows Unit, Okinawa Institute of Science and Technology Graduate University (OIST), 1919-1 Tancha, Onna-son, Okinawa 904-0495, Japan}
\address[kth]{Department of Engineering Mechanics, Royal Institute of Technology (KTH), \\ Stockholm, Sweden}
\address[ntnu]{Department of Energy and Process Engineering, Norwegian University of Science and Technology (NTNU), Trondheim, Norway}
\begin{abstract}
Multiphase, compressible and viscous flows are of crucial importance in a wide range of scientific and engineering problems. Despite the large effort paid in the last decades to develop accurate and efficient numerical techniques to address this kind of problems, current models need to be further improved to address realistic applications. In this context, we propose a numerical approach to the simulation of multiphase, viscous flows where a compressible and an incompressible phase interact in the low-Mach number regime. In this frame, acoustics is neglected but large density variations of the compressible phase can be accounted for as well as heat transfer, convection and diffusion processes. The problem is addressed in a fully Eulerian framework exploiting a low-Mach number asymptotic expansion of the Navier-Stokes equations. A Volume of Fluid approach (VOF) is used to capture the liquid-gas interface, built on top of a massive parallel solver, second order accurate both in time and space. The second-order-pressure term is treated implicitly and the resulting pressure equation is solved with the eigenexpansion method employing a robust and novel formulation. We provide a detailed and complete description of the theoretical approach together with information about the numerical technique and implementation details. Results of benchmarking tests are provided for five different test cases.
\end{abstract}
\end{frontmatter}
%
%\linenumbers
%
% ====================================================================================
%
\section{Introduction}
Multiphase flows of two or more immiscible and viscous fluids are common in a large variety of engineering applications and fundamental scientific problems. Each phase is segregated and gives origin to complex and time-evolving free boundaries where discontinuities in the flow fields exist~\cite{hirt1981vof}. The phases mutually interact exchanging mass, momentum and energy across the free boundaries, the latter undergoing large and complex deformations. It is therefore clear how the description of the problem is extremely challenging both from a theoretical and numerical point of view. Some of the major issues affecting the modeling of multiphase flows arise from the discontinuities in the flow variables and properties across the free boundaries, from the necessity of numerically tracking and reconstructing the interfaces as well as from the need to account for the effect of the surface tension and jump conditions at the interfaces. These aspects are critical in the simulations of incompressible and isothermal multiphase flows, but additional complexity is add when the compressibility needs to be taken into account. In the latter case, heat transfer between the fluid phases and the boundaries must be considered in addition to mutual heat transfer between the phases and density variations. In this paper, the emphasis is on the multiphase flows of two immiscible viscous fluids, one of them being compressible, the other being incompressible. The attention is paid in particular to the low-Mach number flows, where the effect of compressibility is significant and large density variations occur in the compressible phase while acoustics is negligible. This is of great interest for several applications. To mention some, the simulation of bubble-laden flows and boiling flows~\cite{elghobashi2019direct,tanguy2014benchmarks,le2014towards}, as well as the simulation of the fuel jet atomization processes in the combustion chambers of internal combustion engines~\cite{salvador2016numerical,duret2013improving}.\par
A great amount of literature dealing with the numerical simulations of multiphase, viscous fluids has been produced~\cite{picano2015turbulent,balcazar2016coupled,cottet2016semi,ardekani2019turbulent,costa2018effects,rosti2019numerical,dalla2018clustering,rosti2019numerical_a}. Many approaches have been proposed both in the Eulerian and Lagrangian frameworks as well as hybrid methods. Among the latter, the arbitrary Lagrangian Eulerian (ALE) approach~\cite{hirt1974arbitrary,hughes1981lagrangian,ganesan2012arbitrary}. In this frame, interface-conforming grids are used where boundary conditions can be accurately prescribed on the free boundaries of the flow. The main advantage of the ALE methods is the accurate treatment of the interfaces. Nevertheless, the computational cost of this kind of simulations is large due to the adaptive mesh adjustment needed to preserve the conformity of the grid to the time-evolving interfaces.
The need for re-meshing is removed in the frame of the fixed-Eulerian-grid methods. These are hybrid Eulerian-Lagrangian approaches also referred to as front-tracking methods~\cite{unverdi1992front}. The Immersed Boundary Method (IBM) belongs to this class of numerical techniques~\cite{peskin1972flow,breugem2012second,mittal2005immersed}. This approach consists in solving the governing equations for the flow on a fixed Eulerian grid while tracking the free boundaries separating the different phases of the flow by means of Lagrangian markers distributed over the interfaces. An additional forcing is imposed to the fluid, within a neighborhood of the interfaces, such that the boundary conditions are satisfied within a certain degree of accuracy. This class of methods have been successfully applied to the simulation of multi-phase flows~\cite{haeri2012application,kempe2012improved}. Even if fixed-Eulerian-grid methods are computationally more efficient than conforming-grid methods, they suffer from low accuracy in the reconstruction and tracking of the interface.
A popular alternative is the so-called front-capturing methods that are based on a fully Eulerian treatment of the interface tracking and reconstruction. These include essentially the Volume Of Fluid (VOF) method and the Level-Set Method (LSM). The LSM uses a continuous level-set function, usually the signed distance to the interface, to distinguish between the different phases of the flow~\cite{sussman2009stable,sussman1994level,osher1988fronts,gibou2018review,sethian2003level}. Interfaces are accurately defined by an assigned level of the level-set function while the advection of the level-set function itself allows for an accurate tracking of the interfaces. In the LSM framework the interface curvature can be computed easily and accurately, nevertheless, these are not mass-preserving methods. Indeed, the advection of the level-set function may result in a mass loss or gain. On the contrary, the VOF approach, uses a discontinuous colour function to represent each different phase~\cite{hirt1981vof,scardovelli1999direct,gueyffier1999volume,o2018volume}. Interface reconstruction is achieved by a piece-wise representation of the interface in each discrete cell of the computational grid. An accurate computation of the numerical fluxes needed to advect the VOF function is possible and numerical diffusion is avoided making VOF methods extremely accurate from a mass-conservation point of view.  A number of algorithms have been developed to compute the exact interface surfaces from the discrete VOF function. Among these are the Simple Line Interface Calculation (SLIC) method~\cite{noh1976slic}, the Piecewise Linear Interface Calculation (PLIC)~\cite{youngs1982time,youngs1984interface,aulisa2003geometrical}, the Tangent of Hyperbola for Interface Capturing (THINC)~\cite{xiao2005simple} and its multidimensional variant MTHINC~\cite{Ii2012interface}.\par
A major part of the numerical approaches to the simulation of multiphase flows reported by archival literature and referenced above were originally developed for incompressible flows. A great effort has been spent in the last decades to extend these methods to the simulation of compressible multiphase flows~\cite{pelanti2017low,haimovich2017numerical,lunati2006multiscale,saurel2001multiphase,saurel1999multiphase}; nonetheless this is still a very active area of research. Among compressible flows, low-Mach number flows are of great interest for many applications where large density variations occur at low speeds, low subsonic regimes. When addressing the simulation of this flow regime, a big issue arises from the limitation imposed on the time step by the fastest dynamics of the flow. Indeed, in a compressible flow the speed of propagation of pressure waves scales as $1/Ma$, $Ma$ being the Mach number. Many solutions to this problem have been proposed, such as an implicit treatment of the acoustic pressure~\cite{colella1999projection,wall2002semi}. Nevertheless, if the case under examination is dominated by free or forced convection where the amount of energy carried by the acoustics is only a small fraction of the overall energy of the flow, a low-Mach number asymptotic formulation of the Navier-Stokes equations can be used to numerically address the problem. Large density variations can be accounted for, completely neglecting acoustics, but still describing entropy and vorticity modes as well as taking into account compressibility~\cite{muller1998lowmach,majda1985derivation}. In this frame, the pressure is split into two different terms: a zero-order, thermodynamic pressure, $p_0$, and a second-order pressure, $p_2$. The former is governed by the thermodynamic properties of the flow while the latter enters the computation in a similar fashion to that of pressure in incompressible flows~\cite{muller1998lowmach}.\par
In this context, we propose a one-fluid fully Eulerian approach to the numerical simulation of multiphase low-Mach number flows, based on the solution of a low-Mach number asymptotic formulation of the compressible Navier-Stokes equations. For the reconstruction and subsequent advection of the interface between the compressible and incompressible phases, we adopt an algebraic Volume-of-Fluid method (MTHINC~\cite{Ii2012interface}). However, the mathematical and numerical framework can be extended in a straightforward manner to any kind of interface capturing and tracking technique based on the sharp interface approach. The proposed method is implemented in the frame of the pressure-correction methods, taking advantage of a Fast-Fourier-Transform (FFT) based solver for the Poisson equation governing the second-order-pressure of the flow. The effect of the surface tension is accounted for by using the continuum surface force (CSF) model by Brackbill~\cite{brackbill1992continuum}. The implementation is built upon an extensively validated code for the simulation of incompressible flows. The solver uses second order finite difference schemes for space discretization on a fixed Eulerian grid and a second order of accuracy Adams-Bashforth time-marching algorithm~\cite{rosti2017numerical,rosti2018rheology,rosti2019numerical,rosti_ge_jain_dodd_brandt_2019}. We provide a detailed and complete description of the theoretical approach together with information about the numerical technique and implementation details. Because of its numerical efficiency, we believe that this approach is one of the most promising in order to efficiently address the simulation of multiphase, low-Mach number flows, in particular when one of the two phases can be assumed to be incompressible.
%
% ====================================================================================
%
\section{Governing equations}
We consider the flow of two immiscible viscous fluids and let one of them be compressible, the other being incompressible (e.g. a gas-liquid system). Moreover, the compressible phase is assumed to evolve in the low-Mach number regime. A phase indicator function $H$ is defined at position $\mathbf{x}$ and time $t$ in order to distinguish  between the two phases:
%
% EEEEEEEEEEEEEEEEEEEEEEEEEEEEEEEEEEE
\begin{equation}
H(\mathbf{x},t) = \begin{cases}
				1 \hspace{0.5 cm} \text{if $\mathbf{x} \in V_g$}\mathrm{,} \\
				0 \hspace{0.5 cm} \text{if $\mathbf{x} \in V_l$}\mathrm{;} 
			   \end{cases}
\label{eqn:phase_ind}
\end{equation}
% EEEEEEEEEEEEEEEEEEEEEEEEEEEEEEEEEEE
%
where $V_g$ and $V_l$ are the domains pertaining to the gas and liquid phases, divided by a zero-thickness interface $S(t)=V_g\bigcap V_l$. The dynamics of the liquid phase is governed by the standard incompressible Navier-Stokes equations and details are not given here as the framework is well established~\cite{scardovelli1999direct}. On the other hand, a more detailed analysis is presented in the next section for the gas phase.
\subsection{Asymptotic expansions for the compressible phase}
Let now consider the compressible phase only and therefore, unless otherwise stated, all the quantities defined in this section refer to it. In general, if compressibility is taken into account, the gas phase can be described by the following form of the Navier-Stokes and energy equations:
%
% EEEEEEEEEEEEEEEEEEEEEEEEEEEEEEEEEEE
\begin{align}
&\frac{\partial \rho}{\partial t} + \nabla \cdot(\rho \mathbf{u})=0,
\label{eq_3}
\\
&\frac{\partial (\rho \mathbf{u})}{\partial t} + \nabla \cdot (\rho \mathbf{u} \otimes \mathbf{u})=\nabla \cdot \boldsymbol{\tau} - \nabla {p} + \mathbf{f}_{\sigma}+\rho\mathbf{g},
\label{eq_4}
\\
&\frac{\partial(\rho e_t)}{\partial t}+\nabla \cdot (\rho \mathbf{u} e_t) = \nabla \cdot (\boldsymbol{\tau} \cdot \mathbf{u})+\nabla \cdot (k\nabla T) - \nabla \cdot({p\mathbf{u}})+\left(\mathbf{f}_{\sigma}+\rho\mathbf{g}\right)\cdot\mathbf{u},
\label{eq_5}
\\
&p=\rho \mathcal{R} T,
\label{eq_6}
\end{align}
% EEEEEEEEEEEEEEEEEEEEEEEEEEEEEEEEEEE
%
where $\mathbf{u}=(u,v,w)$, $\rho$ and $p$ are the fluid velocity, density and pressure, $\mathbf{g}$ is the gravitational acceleration and $k$ the thermal conductivity. The specific total energy of the flow is denoted as $e_t$ and includes the specific internal energy of the flow, $e$ and the specific kinetic energy: $e_t=e+\mathbf{u}\cdot\mathbf{u}/2$. The Newton-Stokes constitutive relation is assumed such that the viscous stress tensor is
%
% EEEEEEEEEEEEEEEEEEEEEEEEEEEEEEEEEEE
\begin{equation}
\boldsymbol{\tau}= \mu\left[(\nabla\mathbf{u} + \nabla\mathbf{u}^T) - \frac{2}{3}(\nabla \cdot \mathbf{u}) \mathbf{I}\right],
\label{eq_7}
\end{equation}
% EEEEEEEEEEEEEEEEEEEEEEEEEEEEEEEEEEE
%
with $\mathbf{I}$ the identity tensor and $\mu$ the dynamic viscosity. The effect of the surface tension on the interfaces separating the incompressible and compressible phases is modelled by a continuum surface force (CSF)~\cite{brackbill1992continuum}:
%
% EEEEEEEEEEEEEEEEEEEEEEEEEEEEEEEEEEE
\begin{equation}
\mathbf{f}_{\sigma}=\sigma\kappa\delta(\mathbf{x}-\mathbf{x}_s)\mathbf{n},
\label{eq_8}
\end{equation}
% EEEEEEEEEEEEEEEEEEEEEEEEEEEEEEEEEEE
%
where $\sigma$ is the surface tension coefficient, $\kappa$ the curvature of the interface and $\mathbf{n}$ the unit normal on the interface pointing towards the compressible phase. A delta function, $\delta(\mathbf{x}-\mathbf{x}_s)$, is used in Eq.~\eqref{eq_8} to impose the force density, ${\mathbf f}_\sigma$, only at the interface position $\mathbf{x}_s$. The hypothesis of ideal gas is considered, such that the constitutive relation~\eqref{eq_6} holds, where $\mathcal{R}=R/\mathcal{M}$, with $\mathcal{M}$ the molar mass and $R=8.314$ $\mathrm{J/(mol\cdot K)}$ the universal gas constant. The model could be easily improved by using a different constitutive relation, such as Van der Waals equation or more complex cubic equations of state~\cite{battista2014turbulent}. Nonetheless, for the sake of simplicity the discussion is limited to the use of Eq.~\eqref{eq_6}. The hypothesis of calorically perfect gas is assumed too, such that the constant-pressure and constant-volume heat capacities, $c_{p}$ and $c_{v}$, do not depend on the thermodynamic pressure and temperature. Under this hypothesis the energy Eq.~\eqref{eq_5} can be formulated in terms of the sensible internal energy and enthalpy only:
%
% EEEEEEEEEEEEEEEEEEEEEEEEEEEEEEEEEEE
\begin{align}
&e=\Delta h_{T_{ref}}^0+c_v(T-T_{ref})= \Delta h_{T_{ref}}^0+ c_vT,
\\
&h=\Delta h_{T_{ref}}^0+c_p(T-T_{ref})=\Delta h_{T_{ref}}^0+ c_pT,
\end{align}
% EEEEEEEEEEEEEEEEEEEEEEEEEEEEEEEEEEE
%
with $h=e+p/\rho$ the fluid enthalpy, $T$ the temperature and $\Delta h_{T_{ref}}^0$ the fluid enthalpy of formation evaluated at $T_{ref}=0$ $\mathrm{K}$. Eq.~\eqref{eq_3}-\eqref{eq_6} can now be recast in non-dimensional form setting as independent reference scales the density, $\tilde\rho$, the pressure, $\tilde p$, the length, $\tilde L$, the velocity, $\tilde U$ as well as the following derived quantities:
%
% EEEEEEEEEEEEEEEEEEEEEEEEEEEEEEEEEEE
\begin{equation}\nonumber
  \tilde T = \tilde p/(\mathcal{R}\tilde \rho),\ \tilde t=\tilde L/\tilde U,\ \tilde e=\tilde p/ \tilde \rho,\ \tilde f=\tilde \rho\tilde U^2/\tilde L.
\end{equation}
% EEEEEEEEEEEEEEEEEEEEEEEEEEEEEEEEEEE
%
\noindent The scales $\tilde T$, $\tilde t$, $\tilde e$ and $\tilde f$ are the reference temperature, time, specific energy and force per unit volume, respectively. In addition, reference values for the thermal diffusion coefficient, heat capacities, dynamic viscosity, surface tension coefficient of the fluid and gravitational acceleration are $\tilde k,\ \tilde{c}_p,\ \tilde{c}_v,\ \tilde\mu$, $\tilde \sigma$ and $\tilde g$. After some manipulation, the non-dimensional form of Eq.~\eqref{eq_3}~-~\eqref{eq_6} reads:
%
% EEEEEEEEEEEEEEEEEEEEEEEEEEEEEEEEEEE
\begin{align}
&\frac{\partial \rho}{\partial t} + \nabla \cdot(\rho \mathbf{u})=0,
\label{eq_9}
\\
&\frac{\partial (\rho \mathbf{u})}{\partial t} + \nabla \cdot (\rho \mathbf{u} \otimes \mathbf{u})=\frac{1}{Re}\nabla \cdot \boldsymbol{\tau} - \frac{1}{Ma^2}\nabla {p} + \frac{\mathbf{f}_{\sigma}}{We}+\frac{\rho\mathbf{g}}{Fr^2},
\label{eq_10}
\\
&\frac{\partial(\rho e)}{\partial t}+\nabla \cdot (\rho \mathbf{u} h)+Ma^2 \left[\frac{\partial}{\partial t}\left(\rho \frac{\mathbf{u}\cdot\mathbf{u}}{2}\right) + \nabla\cdot\left(\rho \frac{\mathbf{u}\cdot\mathbf{u}}{2} \right)\right] 
\label{eq_a_6}
\\
&\nonumber = \dfrac{\gamma}{\gamma-1}\frac{1}{Re Pr}\nabla \cdot (k\nabla T)+Ma^2\left[\dfrac{1}{Re}\nabla\cdot(\boldsymbol{\tau}\cdot\mathbf{u})+\dfrac{\mathbf{f}_{\sigma}}{We}+\frac{\rho\mathbf{g}}{Fr^2}\right]\cdot\mathbf{u},
\\
&p=\rho T,
\label{eq_11}
\end{align}
% EEEEEEEEEEEEEEEEEEEEEEEEEEEEEEEEEEE<
%
where $Ma=\tilde U/\sqrt{\tilde p/\tilde \rho}$ is a pseudo-Mach number, $Re=\tilde{\rho} \tilde{U} \tilde{L}/\tilde{\mu}$ the Reynolds number, $Pr=\tilde{c}_p\tilde \mu /\tilde k$ the Prandtl number, $We=\tilde \rho \tilde U^2 \tilde L/ \tilde \sigma $ the Weber number and $Fr=\tilde{U}/\sqrt{g\tilde{L}}$ the Froude number, while $\gamma=\tilde c_p/\tilde c_v$. The low-Mach number limit of Eq.~\eqref{eq_9}~-~\eqref{eq_11} can be derived from a single-scale asymptotic perturbation for small Mach numbers~\cite{majda1985derivation}. The pseudo Mach number, $Ma$, is present in all the equations only with the power of two, hence each quantity can be expanded in the following way:
%
% EEEEEEEEEEEEEEEEEEEEEEEEEEEEEEEEEEE
\begin{align}
&\mathbf{f}(\mathbf{x},t)=\mathbf{f}_0(\mathbf{x},t)+\mathbf{f}_2(\mathbf{x},t)Ma^2+O(Ma^3),
\label{eq_1113}
\end{align}
% EEEEEEEEEEEEEEEEEEEEEEEEEEEEEEEEEEE
%
$f$ being a generic vectorial or scalar quantity. It is also possible to proof that the following hold for the product of two scalar quantities:
%
% EEEEEEEEEEEEEEEEEEEEEEEEEEEEEEEEEEE
\begin{align}
&\left[f(\mathbf{x},t)g(\mathbf{x},t)\right]_0=f_0(\mathbf{x},t)g_0(\mathbf{x},t),
\\
&\left[f(\mathbf{x},t)g(\mathbf{x},t)\right]_2=f_2(\mathbf{x},t)g_0(\mathbf{x},t)+f_0(\mathbf{x},t)g_2(\mathbf{x},t).
\end{align}
% EEEEEEEEEEEEEEEEEEEEEEEEEEEEEEEEEEE
%
In the following, we report the procedure to obtain the low-Mach number limit of the momentum Eq.~\eqref{eq_10}, the algebraic manipulation being similar for the remaining ones. Applying~\eqref{eq_1113} to~\eqref{eq_10} leads to:
%
% EEEEEEEEEEEEEEEEEEEEEEEEEEEEEEEEEEE
\begin{align}
&\frac{\partial}{\partial t}\left[(\rho\mathbf{u})_0+(\rho\mathbf{u})_2 Ma^2+O(Ma^3)\right]+
\label{eq_1100}
\\
&\nonumber+\nabla\cdot\left[(\rho \mathbf{u} \otimes \mathbf{u})_0+(\rho \mathbf{u} \otimes \mathbf{u})_2Ma^2+O(Ma^3)\right]=
\\
&\nonumber=\frac{1}{Re}\nabla\cdot\left[{\boldsymbol\tau}_0+{\boldsymbol\tau}_2Ma^2+O(Ma^3)\right]-\frac{1}{Ma^2}\nabla\left[p_0+p_2Ma^2+O(Ma^3)\right]+
\\
&\nonumber+\frac{1}{We}\left[\mathbf{f}_{\sigma0}+\mathbf{f}_{\sigma0}Ma^2+O(Ma^3)\right]+\frac{1}{Fr}\left[(\rho\mathbf{g})_0+(\rho\mathbf{g})_2Ma^2+O(Ma^3)\right].
\end{align}
% EEEEEEEEEEEEEEEEEEEEEEEEEEEEEEEEEEE
%
The terms in Eq.~\eqref{eq_1100} with the same order in $Ma$ can be grouped. After some manipulation, we obtain for the zeroth-order pressure:
%
% EEEEEEEEEEEEEEEEEEEEEEEEEEEEEEEEEEE
\begin{align}
&\nabla p_0=0,
\end{align}
% EEEEEEEEEEEEEEEEEEEEEEEEEEEEEEEEEEE
%
and for the second order correction:
%
% EEEEEEEEEEEEEEEEEEEEEEEEEEEEEEEEEEE
\begin{align}
&\frac{\partial \mathbf{u}_0}{\partial t}+\mathbf{u}_0\cdot \nabla\mathbf{u}_0=\frac{1}{\rho_0}\left[\frac{1}{Re}\nabla \cdot \boldsymbol{\tau}_0 -\nabla p_2+\frac{\mathbf{f}_{\sigma0}}{We}\right]+\frac{\mathbf{g}}{Fr^2}.
\label{eq_1112}
\end{align}
% EEEEEEEEEEEEEEEEEEEEEEEEEEEEEEEEEEE
%
Algebraic manipulation is completely omitted for the continuity and energy equations. The reader is referred to existing references for additional details~\cite{muller1998lowmach,majda1985derivation,meister1999asymptotic}. The final low-Mach number relations are directly given below:
%
% EEEEEEEEEEEEEEEEEEEEEEEEEEEEEEEEEEE
\begin{align}
&\frac{\partial \rho_0}{\partial t} + \nabla \cdot(\rho_0 \mathbf{u}_0)=0,
\label{eq_1110}
\\
&\frac{\partial \mathbf{u}_0}{\partial t}+\mathbf{u}_0\cdot \nabla\mathbf{u}_0=\frac{1}{\rho_0}\left[\frac{1}{Re}\nabla \cdot \boldsymbol{\tau}_0 -\nabla p_2+\frac{\mathbf{f}_{\sigma0}}{We}\right]+\frac{\mathbf{g}}{Fr^2},
\\
&\frac{\partial(\rho_0 e_0)}{\partial t}+\nabla \cdot (\rho_0 \mathbf{u}_0 h_0)=\frac{\gamma}{\gamma-1}\frac{1}{Re Pr}\nabla \cdot (k\nabla T_0),
\\
&p_0=\rho_0 T_0.
\label{eq_1111}
\end{align}
% EEEEEEEEEEEEEEEEEEEEEEEEEEEEEEEEEEE
%
As mentioned above, two different pressure terms appear in these equations: the zeroth-order pressure $p_0$ and the second-order pressure $p_2$. The former, which can be referred to as thermodynamic pressure, is determined by the thermodynamic state of the flow, it is uniform across the spatial field and it is a function of the time only. The latter, conversely, enters the computation in a similar fashion to that of pressure in incompressible flows (e.g. by imposing a prescribed value of the velocity divergence) and it is obtained as solution of a Poisson equation that will be discussed in the following. It is useful to remind that, for the chosen set of reference scales, the non-dimensional sensible enthalpy reads $h_0=\gamma/(\gamma-1)T_0$ while the sensible internal energy $e_0=1/(\gamma-1)T_0$. From here on, the subscript referring to the order of quantities will be omitted except for the pressure terms. After some additional manipulation Eq.~\eqref{eq_1110}~-~\eqref{eq_1111} can be recast as:
%
% EEEEEEEEEEEEEEEEEEEEEEEEEEEEEEEEEEE
\begin{align}
&\frac{\partial \mathbf{u}}{\partial t}+\mathbf{u}\cdot \nabla\mathbf{u}=\frac{1}{\rho}\left[\frac{1}{Re}\nabla \cdot \boldsymbol{\tau} -\nabla p_2+\frac{\mathbf{f}_{\sigma}}{We}\right]+\frac{\mathbf{g}}{Fr^2},
\label{eq_12}
\\
&\frac{\partial T}{\partial t}+\mathbf{u}\cdot \nabla T = \frac{T}{p_0}\left[\frac{1}{Re Pr}\nabla \cdot (k\nabla T)+\frac{\gamma-1}{\gamma}\frac{dp_0}{dt}\right],
\label{eq_13}
\\
&\nabla\cdot\mathbf{u}= \frac{1}{p_0}\left[\frac{1}{Re Pr}\nabla \cdot (k\nabla T)+\frac{1}{\gamma}\frac{dp_0}{dt}\right],
\label{eq_14}
\\
&\frac{dp_0}{dt}=\frac{\gamma}{V}\left[\frac{1}{Re Pr}\int_S k\nabla T\cdot \mathbf{n}\ dS-p_0\int_S\mathbf{u}\cdot\mathbf{n}\ dS\right],
\label{eq_15}
\\
& p_0= \rho T.
\label{eq_16}
\end{align}
% EEEEEEEEEEEEEEEEEEEEEEEEEEEEEEEEEEE
%
\subsection{Final form of the governing equations}
Eq.~\eqref{eq_12},~\eqref{eq_13},~\eqref{eq_14},~\eqref{eq_15} and~\eqref{eq_16} are valid only for the compressible gas phase and can be coupled with those for the incompressible phase using the phase indicator function defined by Eq.~\eqref{eqn:phase_ind}. By doing so, a monolithic system of equations can be obtained describing at the same time the dynamics of both the compressible and incompressible phases. For convenience the system is written in dimensional form,
%
% EEEEEEEEEEEEEEEEEEEEEEEEEEEEEEEEEEE
\begin{align}
&\frac{\partial \mathbf{u}}{\partial t}+\mathbf{u}\cdot \nabla\mathbf{u}=\frac{1}{\rho}\left(\nabla \cdot \boldsymbol{\tau} -\nabla p_2+\mathbf{f}_{\sigma}\right)+\mathbf{g},
\label{eq_17}
\\
&\frac{\partial T}{\partial t}+\mathbf{u}\cdot \nabla T = \frac{1}{\rho c_p}\left[\nabla \cdot (k\nabla T)+\frac{dp_0}{dt}H\right],
\label{eq_18}
\\
&\nabla\cdot\mathbf{u}= \frac{1}{p_0}\left[\frac{\gamma_g-1}{\gamma_g}\nabla \cdot (k\nabla T)-\frac{1}{\gamma_g}\frac{dp_0}{dt}\right]H,
\label{eq_19}
\\
&\frac{dp_0}{dt}=\frac{\gamma_g}{V_g}\left(\frac{\gamma_g-1}{\gamma_g}\int_S k\nabla T\cdot \mathbf{n}\ dS-p_0\int_S\mathbf{u}\cdot\mathbf{n}\ dS\right),
\label{eq_20}
\\
& \rho_g = \dfrac{p_0}{\mathcal{R} T}.
\label{eq_21}
\end{align}
% EEEEEEEEEEEEEEEEEEEEEEEEEEEEEEEEEEE
%
The variables $c_p$, $k$ and $\mu$ are the heat capacity at constant pressure, thermal conductivity and dynamic viscosity of the mixture, respectively. These quantities are averaged, together with the mixture density, using the phase indicator:
%
% EEEEEEEEEEEEEEEEEEEEEEEEEEEEEEEEEEE
\begin{align}
&\rho=\rho_gH+\rho_l(1-H),
\label{eq_22}
\\
&c_p=c_{p,g}H+c_{p,l}(1-H),
\label{eq_23}
\\
&k=k_gH+k_l(1-H),
\label{eq_24}
\\
&\mu=\mu_gH+\mu_l(1-H),
\label{eq_25}
\end{align}
% EEEEEEEEEEEEEEEEEEEEEEEEEEEEEEEEEEE
%
where $c_p$ refers to the specific heat capacity at constant pressure and the subscripts \quotes{$g$} and \quotes{$l$} to the physical parameter of the gas and liquid phases. The ratio $\gamma_g$ depends only on the local thermodynamic variables, but in the present work is taken as a constant and equal to $c_{p,g}/c_{v,g}$. \par
Finally, it is worth remarking that while Eq.~\eqref{eq_17}~-~\eqref{eq_19} hold for both the phases, Eq.~\eqref{eq_20}~-~\eqref{eq_21} are meaningful for the compressible phase only. In particular, Eq.~\eqref{eq_19} reduces to $\nabla \cdot \mathbf{u}=0$ in the liquid regions, where $H=0$. In the monolithic system of equations above, the zeroth-order pressure, $p_0$, results to be constant over each of the closed regions in the computational domain filled by the compressible phase. Its value is set by the constitutive law for ideal gases~\eqref{eq_21} while its temporal rate of change is set by the energy balance provided by Eq.~\eqref{eq_20}. In particular, the surface integrals appearing in Eq.~\eqref{eq_20} have to be intended as integral computed over both the frontier defined by the boundaries of the computational domain and the interface $S$ separating the compressible and incompressible regions. Nevertheless, for its numerical integration, it is more convenient to reformulate Eq.~\eqref{eq_20} in terms of volume integral by means of the divergence theorem:
%
% EEEEEEEEEEEEEEEEEEEEEEEEEEEEEEEEEEE
\begin{align}
  &\frac{dp_0}{dt}=\frac{\gamma_g}{V_g}\left(\frac{\gamma_g-1}{\gamma_g}\int_V H\nabla\cdot(k\nabla T)\ dV-p_0\int_VH\nabla\cdot \mathbf{u}\ dV\right),
  \label{final_p0}
\end{align}
% EEEEEEEEEEEEEEEEEEEEEEEEEEEEEEEEEEE
%
where the volume over which the integrals are computed is the total volume $V=V_g\bigcup V_l$. Eq.~\eqref{final_p0} is derived by employing the change of variable $V_g=HV$, with the differential  $dV_g=HdV$, being $dH=0$ by definition of the phase indicator function~\cite{scardovelli1999direct}.
\section{Numerical methodology}
The numerical solution of Eq.~\eqref{eq_17}-\eqref{eq_21} is addressed on a fixed regular Cartesian grid (e.g. using a uniform and equal spacing, $\Delta x=\Delta y=\Delta z$), with a marker-and-cell arrangement of velocity and pressure points, whereas all scalar fields are defined at the cell centers. Hereafter, we present the numerical discretization of the governing equations following the same order in which they are solved.
\subsection{Interface representation and advection}
The first step of each iteration of the time-marching algorithm consists in the reconstruction of the interface between the two phases and its subsequent advection. As mentioned in the introductory section, we address both the aspects in a fully Eulerian framework using the VOF method to distinguish between each of the flow phases~\cite{hirt1981vof}. By a numerical point of view the indicator function $H$, defined in Eq.~\eqref{eqn:phase_ind}, is updated on the computational grid by the following advection equation:
%
% EEEEEEEEEEEEEEEEEEEEEEEEEEEEEEEEEEE
\begin{equation}
\frac{\partial \Phi}{\partial t}+\nabla\cdot(\mathbf{u} H)=\Phi \nabla \cdot \mathbf{u},
\label{eq_1}
\end{equation}
% EEEEEEEEEEEEEEEEEEEEEEEEEEEEEEEEEEE
%
where the volume fraction, $\Phi$, is defined as the average value of the color function over a discrete computational cell of volume $\Delta V=\Delta x\Delta y\Delta z$:
%
% EEEEEEEEEEEEEEEEEEEEEEEEEEEEEEEEEEE
\begin{equation}
  \Phi=\int_{\Delta V}H(\mathbf{x},t) dV.
  \label{eq_2}
\end{equation}
% EEEEEEEEEEEEEEEEEEEEEEEEEEEEEEEEEEE
%
Coherently with the given definition of $H$ in Eq.~\eqref{eqn:phase_ind}, the volume fraction satisfies $\Phi=1$ in cells occupied by the gas phase only, $\Phi=0$ in that filled only by the liquid and $0<\Phi<1$ in the cells containing the liquid-gas interface. \par
In the present work, we employ as VOF method the multi-dimensional tangent of hyperbola for interface capturing method (MTHINC), originally developed by Ii et al.~\cite{Ii2012interface} \scapin{and more recently applied to complex flows cases both in laminar~\cite{rosti2019numerical,de-vita_rosti_caserta_brandt_2020a,de-vita_rosti_caserta_brandt_2019a} and in turbulent conditions~\cite{rosti_ge_jain_dodd_brandt_2019}}. The description of the procedure is not reported here as our procedure is not limited to a specific interface capturing/tracking method; additional details are provided in the references~\cite{Ii2012interface,rosti2019numerical}. Once the interface is reconstructed, the advection step is performed using the standard directional splitting approach, originally developed by Puckett~et~al.~\cite{puckett1997high} and Aulisa~et~al.~\cite{aulisa2003geometrical}. An additional correction is used in order to account for the non-zero divergence of the interface velocity~\cite{scapin2020volume}, that in the present framework coincides with the one-fluid velocity $\mathbf{u}$. Finally, the thermodynamic properties independent of temperature ($\mu$, $k$ and $c_p$) are updated using the relations~\eqref{eq_23},~\eqref{eq_24} and~\eqref{eq_25}.
\subsection{Thermodynamic pressure and temperature equation}
The next step consists in the computation of the updated thermodynamic pressure $p_0^{n+1}$ and temperature $T^{n+1}$ using a second order Adams-Bashforth time-marching algorithm:
%
% EEEEEEEEEEEEEEEEEEEEEEEEEEEEEEEEEEE
\begin{align}
	&p_0^{n+1}=p_0^n+\Delta t^{n+1}\left[\left(1+\dfrac{1}{2}\dfrac{\Delta t^{n+1}}{\Delta t^n}\right)RP^n-\left(\dfrac{1}{2}\dfrac{\Delta t^{n+1}}{\Delta t^n}\right)RP^{n-1}\right],
\label{eq_26}
\\
	&T^{n+1}=T^n+\Delta t^{n+1}\left[\left(1+\dfrac{1}{2}\dfrac{\Delta t^{n+1}}{\Delta t^n}\right)RT^n-\left(\dfrac{1}{2}\dfrac{\Delta t^{n+1}}{\Delta t^n}\right)RT^{n-1}\right].
\label{eq_27}
\end{align}
% EEEEEEEEEEEEEEEEEEEEEEEEEEEEEEEEEEE
%
The variables $\Delta t^{n+1}$ and $\Delta t^{n}$ represent the time step evaluated at the time levels $n+1$ and $n$, respectively. The time step is chosen to fulfill the temporal stability requirements as explained in section~\ref{sec:time_step}. The terms $RP$ and $RT$ are the right-hand side of the thermodynamic pressure and temperature, provided below in a semi-discrete notation:
%
% EEEEEEEEEEEEEEEEEEEEEEEEEEEEEEEEEEE
\begin{align}
	&RP^n=\frac{\gamma_g}{V_g^n}\left[\frac{\gamma_g-1}{\gamma_g}\int_{V^n}\Phi^{n}\nabla\cdot(k^n\nabla T^n)\ dV-p_0^n\int_{V^n}\Phi^{n}\nabla\cdot\mathbf{u}^n\ dV\right],
\label{eq_30}
\\
	&RT^n=-\mathbf{u}^n\cdot \nabla T^n +\frac{1}{\rho^n c_p^{n+1}}\left[\nabla \cdot (k^{n+1}\nabla T^n)+\left(\dfrac{dp_0}{dt}\right)^n \Phi^n\right],
\label{eq_31}
\end{align}
% EEEEEEEEEEEEEEEEEEEEEEEEEEEEEEEEEEE
%
where the rate of change of the thermodynamic pressure is computed as $(dp_0/dt)^n=RP^n$. The gas volume $V_g^n$ over which Eq.~\eqref{eq_26} is integrated can be approximated as:
%
% EEEEEEEEEEEEEEEEEEEEEEEEEEEEEEEEEEE
\begin{equation}
 V_g^n = \sum_{i=1}^{N_x}\sum_{j=1}^{N_y}\sum_{k=1}^{N_z}\Phi_{i,j,k}^{n}\Delta x\Delta y\Delta z, 
 \label{eqn:v_gas}
\end{equation}
% EEEEEEEEEEEEEEEEEEEEEEEEEEEEEEEEEEE
%
$N_x$, $N_y$ and $N_z$ being the number of grid points along the $x$, $y$ and $z$ directions, respectively. All the spatial terms in Eq.~\eqref{eq_30}~and~\eqref{eq_31} are discretized by second order central schemes, except for the temperature convection. The discretization of this latter is based on the $5^{th}$-order WENO5 scheme described in reference~\cite{castro2011high}. Once $T^{n+1}$ and $p_0^{n+1}$ are known, the volumetric density field is updated using Eq.~\eqref{eq_25} while the gas density is computed using the equation of state, e.g. $\rho_g^{n+1}=p_{0}^{n+1}/\mathcal{R}T^{n+1}$. It should be noted that, following this approach, the right-hand side the energy equation, $RT^{n+1}$, contains the specific heat capacity, $c_p$, evaluated at the new time level, $n+1$, while the density, $\rho$, is only available at the previous time level, $n$. Despite this inconsistency, dictated by the coupling between the density and temperature field, the approach adopted in the present work provides convergent and stable results in all the validation cases discussed in the following. 
This inconsistency could be fixed by computing $\rho^{n+1}$ via $\Phi^{n+1}$, but employing the old value of $p_0^n$ and $T^n$. Even if we do not provide the numerical results, we tested this alternative solution observing that the numerical results of the two approaches were indistinguishable.
\subsection{Flow solver}\label{sec:flow_solver}
In order to impose that the velocity field $\mathbf{u}^{n+1}$ satisfies the divergence constraint given by Eq.~\eqref{eq_34}, a pressure-correction scheme based on the Adams-Bashforth method is employed and summarized below in semi-discrete notation:
%
% EEEEEEEEEEEEEEEEEEEEEEEEEEEEEEEEEEE
\begin{align}
     &\mathbf{u}^*=\mathbf{u}^{n}+\Delta t^{n+1}\left[\left(1+\dfrac{1}{2}\dfrac{\Delta t^{n+1}}{\Delta t^n}\right)\mathbf{RU}^n-\left(\dfrac{1}{2}\dfrac{\Delta t^{n+1}}{\Delta t^n}\right)\mathbf{RU}^{n-1}\right],\label{eq_29} \\
     &\nabla\cdot\left(\dfrac{1}{\rho^{n+1}}\nabla p_2^{n+1}\right)=\dfrac{1}{\Delta t^{n+1}}\left[\nabla\cdot\mathbf{u}^{*}-\nabla\cdot\mathbf{u}^{n+1}\right],\label{eqn:lap_p} \\
     &\mathbf{u}^{n+1} = \mathbf{u}^*-\dfrac{\Delta t^{n+1}}{\rho^{n+1}}\nabla p_2^{n+1}\label{eqn:corr},
\end{align}
where $\mathbf{u}^*$ is the predicted velocity. The right-hand side $\mathbf{RU}^n$ is computed as:
\begin{align}
&\mathbf{RU}^n=-\mathbf{u}^n\cdot \nabla\mathbf{u}^n+\dfrac{1}{\rho^{n+1}}\left[\nabla\cdot\boldsymbol{\tau}(\mu^{n+1},\mathbf{u}^n)+\mathbf{f}_{\sigma}^{n+1}+\rho^{n+1}\mathbf{g}\right]\mathrm{,}
\label{eq_32}
\end{align}
% EEEEEEEEEEEEEEEEEEEEEEEEEEEEEEEEEEE
%
where both the convection and diffusion terms are discretized by central schemes. More specifically, the former is discretized in divergence form as $\nabla\cdot(\mathbf{u}\mathbf{u})-\mathbf{u}\nabla\cdot\mathbf{u}$ whereas the latter is treated in a fully conservative form. The color function gradients, $\nabla\Phi^{n+1}$, in the expression for the surface tension force, $\mathbf{f}_{\sigma}$, are estimated with the Youngs' method~\cite{youngs1982time,youngs1984interface}. Finally, the divergence constraint in Eq.~\eqref{eqn:lap_p} is computed directly as
%
% EEEEEEEEEEEEEEEEEEEEEEEEEEEEEEEEEEE
\begin{align}
	&\nabla\cdot\mathbf{u}^{n+1}= \frac{1}{p_0^{n+1}}\left[\frac{\gamma_g-1}{\gamma_g}\nabla \cdot (k^{n+1}\nabla T^{n+1})-\frac{1}{\gamma_g}\left(\dfrac{dp_0}{dt}\right)^{n+1}\right]\Phi^{n+1}\mathrm{.}
\label{eq_34}
\end{align}
% EEEEEEEEEEEEEEEEEEEEEEEEEEEEEEEEEEE
%
It should be remarked, that the evaluation of the term $(dp_0/dt)^{n+1}$ in Eq.~\eqref{eq_34} implies the use of $\mathbf{u}^{n+1}$, which is unknown. A possible strategy is to introduce inside each time-step a further inner loop to couple $p_0^{n+1}$ and $\mathbf{u}^{n+1}$~\cite{daru2010numerical}. In the present work, we adopt a simpler but equally robust solution based on an extrapolation in time of the term $(dp_0/dt)^{n+1}$:
%
% EEEEEEEEEEEEEEEEEEEEEEEEEEEEEEEEEEE
\begin{equation} 
  \left(\dfrac{dp_0}{dt}\right)^{n+1}\simeq \left(1+\dfrac{\Delta t^{n+1}}{\Delta t^n}\right)\left(\dfrac{dp_0}{dt}\right)^{n}-\left(\dfrac{\Delta t^{n+1}}{\Delta t^n}\right)\left(\dfrac{dp_0}{dt}\right)^{n-1},
\end{equation}
% EEEEEEEEEEEEEEEEEEEEEEEEEEEEEEEEEEE
%
where $(dp_0/dt)^{n}=RP^n$ and $(dp_0/dt)^{n-1}=RP^{n-1}$ computed using Eq.~\eqref{eq_30}. The resulting system of equations~\eqref{eq_29},~\eqref{eqn:lap_p} and~\eqref{eqn:corr}, is formally identical to that used by any incompressible two-fluid solver where the one-fluid velocity field is not divergence-free. In general, this may occur in presence of phase change at the interface (see reference~\cite{tanguy2014benchmarks}), but in our mathematical framework it can occur only in the regions occupied by the compressible phase. 
\subsubsection{Pressure equation}
A key feature of any two-fluid solver is the ability to impose accurately and efficiently the divergence constraint on the velocity field, this task being directly related to the numerical procedure used to solve the pressure equation, Eq.~\eqref{eqn:lap_p}. A possible approach is based on the use of iterative multigrid solvers. Despite the success and the widespread use of these solvers, the solution is not exact, but satisfied up to a controlled tolerance, usually of the order $\varepsilon=10^{-7}$~-~$10^{-8}$. Moreover, since the coefficients of the Poisson equation vary in space, the system matrix must be recomputed at each time-step. Alternatively, when the pressure boundary conditions are homogeneous~\cite{costa2018fft}, a possible solution is to transform Eq.~\eqref{eqn:lap_p} into a constant coefficient equation and apply the method of the eigenexpansion~\cite{wilhelmson1977direct,schumann1988fast} to solve the pressure equation exactly with spectral accuracy. The resulting pressure equation is still to be solved in an iterative manner starting with an initial guess. Different methods based on the latter approach are available in literature, the main difference being how the variable coefficient pressure equation is recast into a constant coefficient problem. In the following, we review two of these methods, that have been recently proposed and designed to efficiently solve in an iterative manner the Poisson equation with the method of eigenexpansion. It should be noted that, these methods have been successfully implemented in numerical codes that share with our one a similar parallelization strategy based on the~\textbf{2DECOMP\&FFT}~\cite{li20102decomp} library. Finally, we will introduce a new methodology that proves to be more efficient and suitable for two-phase flows with capillary effects and sharp gradients between the two phases.
\begin{itemize}
\item Method I: this method has been proposed by Motheau and Abraham~\cite{motheau2016high}. It is designed for low-Mach number reactive flows, aiming at decreasing the number of iterations of the previously developed FFT-based solvers for combustion applications. The methodology consists in a semi-implicit approach that first requires an iterative procedure to solve the following constant coefficient Poisson equation:
%
% EEEEEEEEEEEEEEEEEEEEEEEEEEEEEEEEEEE
\begin{equation}
\nabla^2 p_2^{s+1} = \nabla\cdot\left[\left(1-\dfrac{\tilde{\rho}_0^{n+1}}{\rho^{n+1}}\right)\nabla p_2^{s}\right]+\dfrac{\tilde{\rho}_0^{n+1}}{\Delta t^{n+1}}\left(\nabla\cdot\mathbf{u}^*-\nabla\cdot\mathbf{u}^{n+1}\right)\mathrm{,}
\label{eqn:poisson_me1}
\end{equation} 
% EEEEEEEEEEEEEEEEEEEEEEEEEEEEEEEEEEE
%
where, $p_2^{s+1}$ and $p_2^{s}$ are the hydrodynamic pressure at two subsequent iterations and $\tilde{\rho}_0^{n+1}$ is the minimum value of $\rho^{n+1}$ over the computational domain. After the iterative loop to compute $p_2^{n+1}$, a modified correction step is applied:
%
% EEEEEEEEEEEEEEEEEEEEEEEEEEEEEEEEEEE
\begin{equation}
\mathbf{u}^{n+1} = \mathbf{u}^*-\Delta t^{n+1}\left[\dfrac{1}{\tilde{\rho}_0^{n+1}}\nabla p_2^{n+1}+\left(\dfrac{1}{\rho^{n+1}}-\dfrac{1}{\tilde{\rho}_0^{n+1}}\right)\nabla p_2^{n+1,q}\right]\mathrm{,}
\label{eqn:poisson_me2}
\end{equation} 
% EEEEEEEEEEEEEEEEEEEEEEEEEEEEEEEEEEE
%
where $p_2^{n+1,q}$ is the second-to-last hydrodynamic pressure of the iterative procedure at the new time-level. The main advantage of this method is the ability to effectively impose the velocity divergence up to machine accuracy, by setting a residual threshold to solve Eq.~\eqref{eqn:poisson_me1} to $\varepsilon_t = 10^{-6}-10^{-8}$, being the residual $\varepsilon=||p_2^{s+1}-p_2^{s}||$). Nevertheless, we find that using this approach in the case of a rising bubble (see section~\ref{sec:rising_bubble}), the number of iterations required to achieve convergence is of the order of one hundred.
\item Method II: this approach, proposed by Bartholomew and Laizet~\cite{bartholomew2019new} and designed for non-Boussinesq gravity currents, is based on the rearrangement of Eq.~\eqref{eqn:lap_p} as a constant coefficient Poisson equation:
%
% EEEEEEEEEEEEEEEEEEEEEEEEEEEEEEEEEEE
\begin{equation}
\nabla^2 p_2^{s+1} = \nabla^2 p_2^{s} + \tilde{\rho}\left[\dfrac{1}{\Delta t^{n+1}}\left(\nabla\cdot\mathbf{u}^*-\nabla\cdot\mathbf{u}^{n+1}-\nabla\cdot\dfrac{1}{\rho^{n+1}}\nabla p_2^s\right)\right]\mathrm{,}
\label{eqn:poisson_laizet}
\end{equation}
% EEEEEEEEEEEEEEEEEEEEEEEEEEEEEEEEEEE
%
where $p_2^{s+1}$ and $p_2^{s}$ are the hydrodynamic pressure at two subsequent iterations. Eq.~\eqref{eqn:poisson_laizet} is solved in an iterative manner until convergence. After that, the correction step~\eqref{eqn:corr} is performed to obtain the new velocity field $\mathbf{u}^{n+1}$. The modified density $\tilde{\rho}$ is taken as the harmonic mean between $\rho_l$ and $\rho_g$, as suggested by the authors to improve convergence. To satisfy the divergence constraint up to machine accuracy, the threshold residual should be set to $\varepsilon_t=10^{-12}$. In section~(\ref{sec:rising_bubble}) we will show that this approach requires a lower number of iterations than the previous one, but still higher than the one we are going to present next.
\item Method III: the basic idea behind this third approach, proposed here, is to rearrange the Poisson equation into a constant-coefficient form  by employing the correction step~\eqref{eqn:corr}:
%
% EEEEEEEEEEEEEEEEEEEEEEEEEEEEEEEEEEE
\begin{equation}
\begin{split}
\dfrac{1}{\rho^{n+1}}\nabla^2p_2^{n+1} + \nabla\left(\dfrac{1}{\rho^{n+1}}\right)\cdot\nabla p_2^{n+1} &= \dfrac{1}{\Delta t^{n+1}}\left(\nabla\cdot\mathbf{u}^*-\nabla\cdot\mathbf{u}^{n+1}\right)\mathrm{,} \\
\nabla^2 p_2^{n+1}-\dfrac{1}{\rho^{n+1}}\nabla\rho^{n+1}\cdot\nabla p_2^{n+1} &= \dfrac{\rho^{n+1}}{\Delta t^{n+1}}\left(\nabla\cdot\mathbf{u}^*-\nabla\cdot\mathbf{u}^{n+1}\right)\mathrm{.}
\end{split}
\label{eqn:transfor}
\end{equation}
% EEEEEEEEEEEEEEEEEEEEEEEEEEEEEEEEEEE
%
Using the vector calculus identity $\rho\nabla\cdot\mathbf{u}=\nabla\cdot(\rho\mathbf{u})-\mathbf{u}\cdot\nabla\rho$, we finally rewrite Eq.~\eqref{eqn:transfor} as:
%
% EEEEEEEEEEEEEEEEEEEEEEEEEEEEEEEEEEE
\begin{align}
\nabla^2 p_2^{n+1} = \dfrac{1}{\Delta t^{n+1}}\left[\nabla\cdot(\rho^{n+1}\mathbf{u}^*)-\rho^{n+1}\nabla\cdot\mathbf{u}^{n+1}-\mathbf{u}^{n+1}\cdot\nabla\rho^{n+1}\right]\label{eqn:new_poisson}\\+\nonumber\dfrac{1}{\Delta t^{n+1}}\underbrace{\left[\left(\mathbf{u}^{n+1}-\mathbf{u}^*+\dfrac{\Delta t^{n+1}}{\rho^{n+1}}\nabla p_2^{n+1}\right)\cdot\nabla\rho^{n+1}\right]}_{\text{$=0$ due to Eq.~\eqref{eqn:corr}}}\mathrm{.}
\end{align}
% EEEEEEEEEEEEEEEEEEEEEEEEEEEEEEEEEEE
%
As $p_{2}^{n+1}$ and $\mathbf{u}^{n+1}$ are both unknown, we solve Eq.~\eqref{eqn:new_poisson} together with Eq.~\eqref{eqn:corr} by an iterative loop, as reported in the pseudocode~\ref{algo_iii}. Two interesting features emerge when using this method. First, the constant-coefficient Poisson Eq.~\eqref{eqn:new_poisson} is an equivalent and exact formulation of its variable counterpart~\eqref{eqn:lap_p}, derived using the correction step~\eqref{eqn:corr}. This represents a major difference with respect to the previous two methods, Eq.~\eqref{eqn:poisson_me1}~and~\eqref{eqn:poisson_laizet}, which are only a consistent but not exact recast of Eq.~\eqref{eqn:lap_p}. Second, this method allows us to define and control the residual of the iterative procedure on the basis of the velocity divergence, which represents the constraint to be imposed on the flow field. These two advantages come at the cost of performing a correction step for each iteration of the loop. Nevertheless, the additional computational cost is more than compensated by the lower numbers of iterations required to achieve convergence as the solution of the Poisson equation is often the most expensive part in standard two-fluid solvers. As we will show in the result section, this approach requires a significantly lower number of iterations.
\end{itemize}
\begin{algorithm}
\caption{Solution of the pressure equation with Method III}
\begin{algorithmic}[1]
\STATE $s = 0,$
\STATE $\mathbf{u}^s=\mathbf{u}^*,$
\STATE $\varepsilon=a\varepsilon_t$ with $a>1.$
\WHILE{$\varepsilon>\varepsilon_t$}
        \STATE $s=s+1\mathrm{,}$
	\STATE $\nabla^2p_2^{s+1}=\frac{1}{\Delta t^{n+1}}\left[\nabla\cdot\left(\rho^{n+1}\mathbf{u}^*\right)-\rho^{n+1}\nabla\cdot\mathbf{u}^{n+1}-\mathbf{u}^s\cdot\nabla\rho^{n+1}\right]\mathrm{,}$
        \STATE $\mathbf{u}^{s+1}=\mathbf{u}^*-\frac{\Delta t^{n+1}}{\rho^{n+1}}\nabla p_2^{s+1}\mathrm{,}$
        \STATE $\varepsilon=\displaystyle{\sum_{i=1}^{N_x}\sum_{j=1}^{N_y}\sum_{k=1}^{N_z}} ||\nabla\cdot\mathbf{u}_{i,j,k}^{s+1}-\nabla\cdot\mathbf{u}_{i,j,k}^{n+1}||\mathrm{.}$
\ENDWHILE
\STATE $p_2^{n+1}=p_2^{s+1},$
\STATE $\mathbf{u}^{n+1}=\mathbf{u}^{s+1}.$
\end{algorithmic}
\label{algo_iii}
\end{algorithm}
The proposed methodology is outlined in the pseudocode~\ref{algo_iii} where the iterations are performed until $\varepsilon \le \varepsilon_{t}$, is satisfied. The residual $\varepsilon$ is computed with a summation over the whole computational domain whereas the threshold value $\varepsilon_t$ is chosen considering the trade-off between the number of iterations required to achieve convergence and the minimization of the residual error. Unless otherwise stated, in the present work, $\varepsilon_t$ equal to $10^{-14}$ has been set in order to impose the divergence constraint on the final velocity field, $\mathbf{u}^{n+1}$, with machine accuracy. Finally, it should be noted that the first two terms of Eq.~\eqref{eqn:new_poisson} do not vary over the solution cycle and can be pre-computed just before the iterative procedure to reduce the execution time, whereas the last term, $\mathbf{u}^{s}\cdot\nabla\rho^{n+1}$ needs to be updated at every iteration of the pressure-correction loop. \par
Before proceeding to the discussion of the validation cases, it is worth remarking that the proposed mathematical and numerical framework can be naturally extended to any interface capturing and tracking method consistent with the sharp-interface definition of the phase indicator function as in Eq.~\eqref{eqn:phase_ind}, e.g. Volume-of-Fluid, Level-set and Front-Tracking methods. Furthermore, as also in different implementations of the diffuse interface approach (e.g. phase field models based on the Cahn-Hilliard and Cahn-Allen equations) the numerical procedure is typically based on the solution of a variable coefficient Poisson equation, we believe that the methodology proposed here can be helpful also to generalize the phase field theory in a low-Mach number framework.
\subsection{Time step restriction}\label{sec:time_step}
The time step $\Delta t^{n+1}$ is estimated from the stability constraints of the overall system:
%
% EEEEEEEEEEEEEEEEEEEEEEEEEEEEEEEEEEE
\begin{equation}
	\Delta t^{n+1}=C_{\Delta t}\min(\Delta t_c,\Delta t_{\sigma},\Delta t_{\mu},\Delta t_e)^{n+1}\mathrm{,}
	\label{eqn:max_dt}
\end{equation}
% EEEEEEEEEEEEEEEEEEEEEEEEEEEEEEEEEEE
%
where $\Delta t_c$, $\Delta t_{\sigma}$, $\Delta t_{\mu}$ and $\Delta t_e$ are the maximum allowable time steps due to convection, surface tension, momentum diffusion and thermal energy diffusion, respectively. These can be determined as suggested in reference~\cite{kang2000boundary}:
%
% EEEEEEEEEEEEEEEEEEEEEEEEEEEEEEEEEEE
\begin{equation}
	\begin{aligned}
	\Delta t_c & =\left(\dfrac{|u_{x,\max}|}{\Delta x}+\dfrac{|u_{y,\max}|}{\Delta y}+\dfrac{|u_{z,\max}|}{\Delta z}\right)^{-1}\mathrm{,} \\
	\Delta t_{\mu}&=\left[\max\left(\dfrac{\mu_g}{\tilde{\rho}_g^m},\dfrac{\mu_l}{\rho_l}\right)\left(\dfrac{2}{\Delta x^2}+\dfrac{2}{\Delta y^2}+\dfrac{2}{\Delta z^2}\right)\right]^{-1}\mathrm{,} \\
	\Delta t_{\sigma}&=\sqrt{\dfrac{(\tilde{\rho}_g^m+\rho_l)\min(\Delta x,\Delta y,\Delta z)^3}{4\pi\sigma}}\mathrm{,} \\
	\Delta t_{e}&=\left[\max\left(\dfrac{k_g}{\tilde{\rho}_g^m},\dfrac{k_l}{\rho_lc_{p,l}}\right)\left(\dfrac{2}{\Delta x^2}+\dfrac{2}{\Delta y^2}+\dfrac{2}{\Delta z^2}\right)\right]^{-1}\mathrm{,}
	\end{aligned}
	\label{eqn:diff_con_t}
\end{equation}
% EEEEEEEEEEEEEEEEEEEEEEEEEEEEEEEEEEE
%
where $|u_{i,\max}|$ is an estimate of the maximum value of the $i^{th}$ component of the flow velocity and $\tilde{\rho}_g^m$ is the minimum gas density computed over the computational domain to account for the compressible effects. For the cases presented here, $C_{\Delta t}=0.25$ was found sufficient for a stable and accurate time integration and, unless otherwise stated, this value has been employed for the validation cases.\par
%
% ====================================================================================
%
\section{Validation and testing}
In order to validate and test the proposed numerical approach, five different flow configurations are considered, denoted \textit{C1a} and \textit{C1b} \textit{C2}, \textit{C3} and \textit{C4}. The first two simulations, \textit{C1a} and \textit{C1b}, reproduce the two-dimensional flow originating in a fluid system made of alternating gaseous and liquid bands at different initial densities and temperatures confined in a periodic, free-slip channel. We believe that these test cases are particularly significant to highlight the capabilities of the proposed numerical methodology. The third simulation, \textit{C2}, reproduces a two-dimensional gas bubble rising in an incompressible liquid medium and is used as a quantitative validation against a reference case from archival literature. We select this test case in order to perform the comparison among the different methods analyzed in section~\ref{sec:flow_solver}. The previous setup is also used to study the flow in the presence of three rising bubbles, case $\textit{C3}$. The last simulation, \textit{C4}, reproduces a time-evolving, plane mixing layer originating  between two streams  at different temperatures and opposite velocities. One of the streams is assumed to be compressible, the other being incompressible. The effect of the temperature gradients on the temporal evolution of the mixing layer is fully described by means of the low-Mach number asymptotic approach, taking into consideration thermal diffusion as well as density gradients in the flow.
%
% TTTTTTTTTTTTTTTTTTTTTTTTTTTTTTTTTTTTTT
\begin{table}[t!]
\centering
\begin{tabular}{lcccccccc}
\hline
& $\mu_l/\mu_g$ & $k_l/k_g$ & $c_{v}/c_{v,g}$ & $c_p/c_{p,g}$ & $\tilde{\sigma}$ & $\tilde{\mathcal{R}}$ & $\tilde{p}$ & $\tilde{g}$ \\
\hline
&- & - & - & - & $\mathrm{N/m}$ & $\mathrm{J/(kg\cdot K)}$ & $\mathrm{Pa}$ & $\mathrm{m/s^2}$ \\
\hline
\textit{\ref{sec:case1}}& $1$ & $1$ & $1$ & $1$ & $24.5$ & $300$ & $9\cdot 10^4$ & 0.0 \\
\hline
\textit{\ref{sec:case2}}& $10$ & $20$ & $5.76$ & $4.18$ & $24.5$ & $300$ & $9\cdot 10^4$ & 0.98 \\
\hline
\textit{\ref{sec:case3}}& $10$ & $20$ & $5.76$ & $4.18$ & $24.5$ & $300$ & $9\cdot 10^4$ & 0.98 \\
\hline
\textit{\ref{sec:case4}}& $1$ & $1$ & $1$ & $1$ & $0$ & $300$ & $9\cdot 10^4$ & 0 \\
\hline
\end{tabular}
\caption{Physical parameters of the fluids for cases \textit{C1a}, \textit{C1b}, \textit{C2}, \textit{C3} and \textit{C4}: viscosity ratio, $\mu_l/\mu_g$, thermal conductivity ratio, $k_l/k_g$, specific heat capacity ratios at constant pressure, $c_{p}/c_{p,g}$ and constant volume $c_{v}/c_{v,g}$, reference surface tension coefficient, $\tilde \sigma$, reference specific constant of the gas phase, $\tilde{\mathcal{R}}$, reference thermodynamic pressure, $\tilde p$, and reference gravitational acceleration, $\tilde g$. The subscript, \quotes{$g$}, refers to the gaseous phase and the subscript \quotes{$l$} to the liquid.}
\label{tab_param}
\end{table}
% TTTTTTTTTTTTTTTTTTTTTTTTTTTTTTTTTTTTTT
%
\subsection{Expansion of gas bands enclosed by an incompressible medium}\label{sec:case1}
The test case \textit{C1} reproduces the two-dimensional, isochoric (\textit{C1a}) and isobaric (\textit{C1b}) transformation of a compressible gas band enclosed within an incompressible liquid medium. All the quantities are provided in the non-dimensional frame, the reference values and the simulation parameter being reported in table~\ref{tab_param}. The domain is rectangular and extends, in non-dimensional units, for $L_x/{\tilde L}=4$ and $L_y/{\tilde L}=0.5$, ${\tilde L}$ being the reference length scale. The domain is discretized using $N_x\times N_y= 128 \times 16$ nodes. Free-slip boundary conditions are applied to the lower and upper edges of the computational domain while an adiabatic, zero-gradient boundary condition is prescribed to the temperature equation. A periodic boundary condition is applied along the $x$ direction. The isochoric case, \textit{C1a}, considers a rectangular gas band of width $b/{\tilde L}=1$ that splits the domain into two parts filled by an incompressible liquid. The band is initially centred around the axial position $x_c/{\tilde L}=3$ and extends over the whole domain in the $y$ direction. 
%
% FFFFFFFFFFFFFFFFFFFFFFFFFFFFFFFFFFFFFF
\begin{figure}[ht!]
\centering
\subfigure[]{\includegraphics[width=0.75\textwidth]{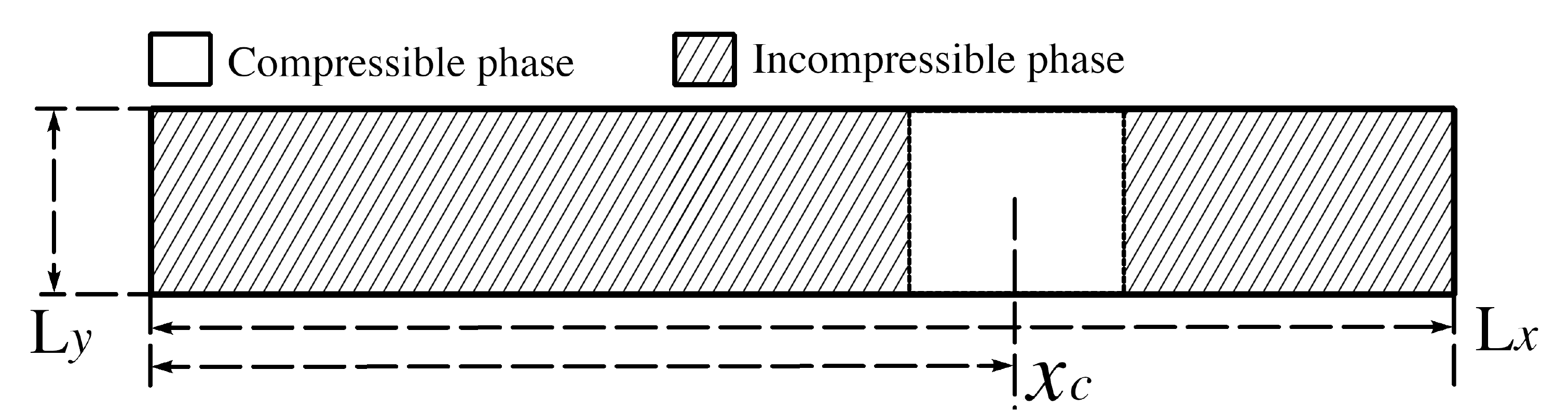}\label{fig_dmn_c1}}
\subfigure[]{\includegraphics[width=0.48\textwidth]{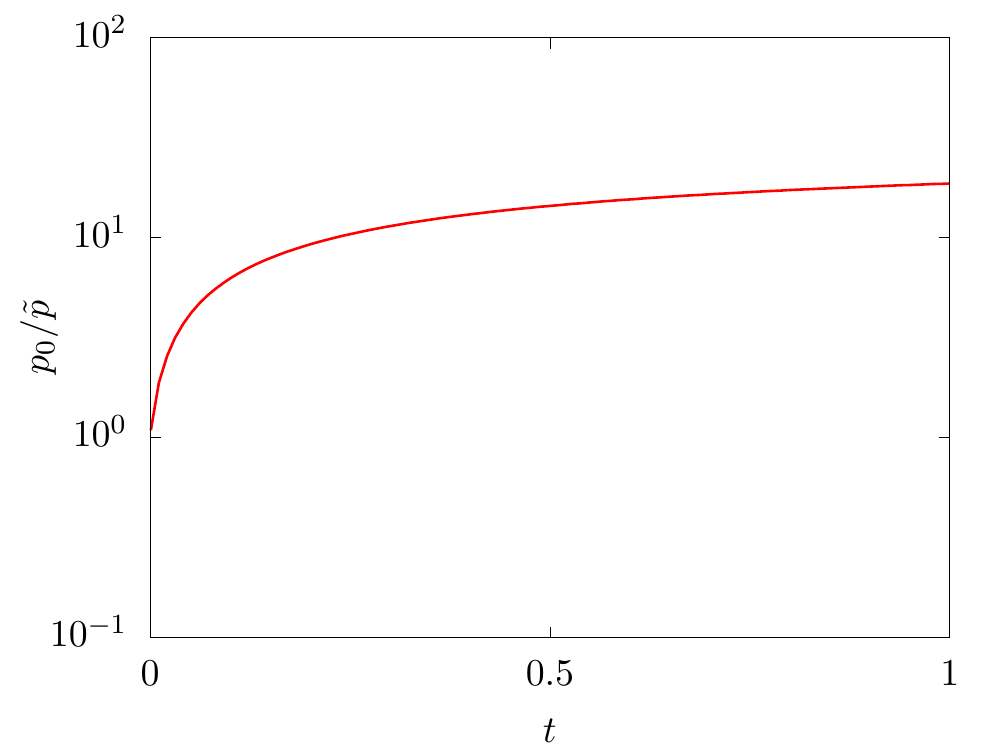}\label{fig_c1_a}}
\subfigure[]{\includegraphics[width=0.48\textwidth]{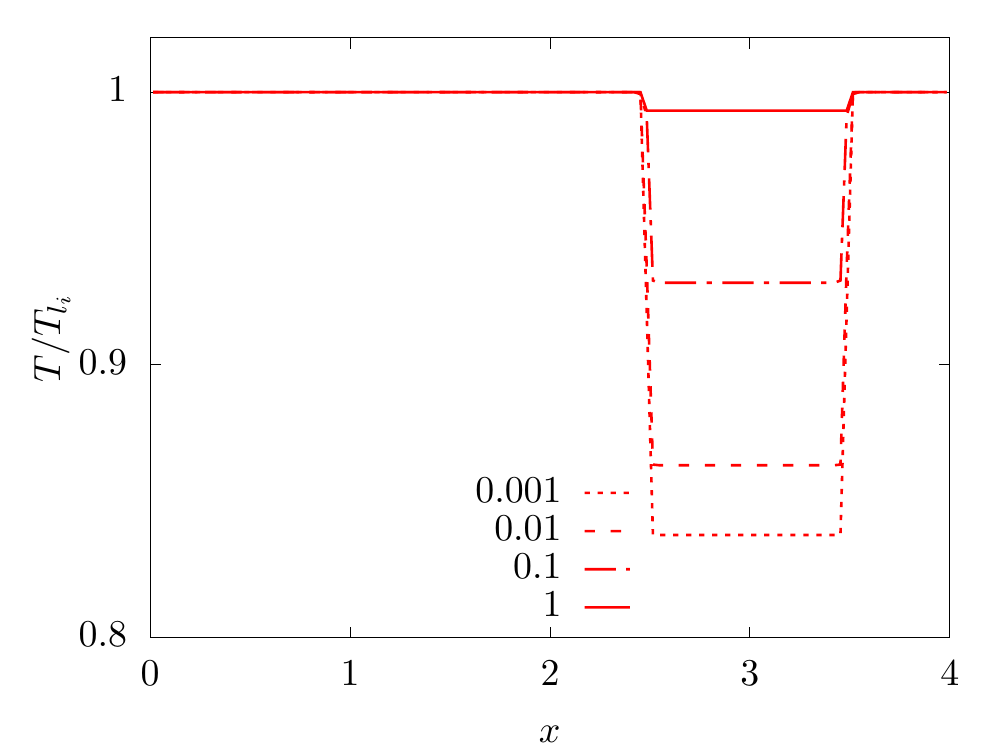}\label{fig_c1_b}}
\subfigure[]{\includegraphics[width=0.48\textwidth]{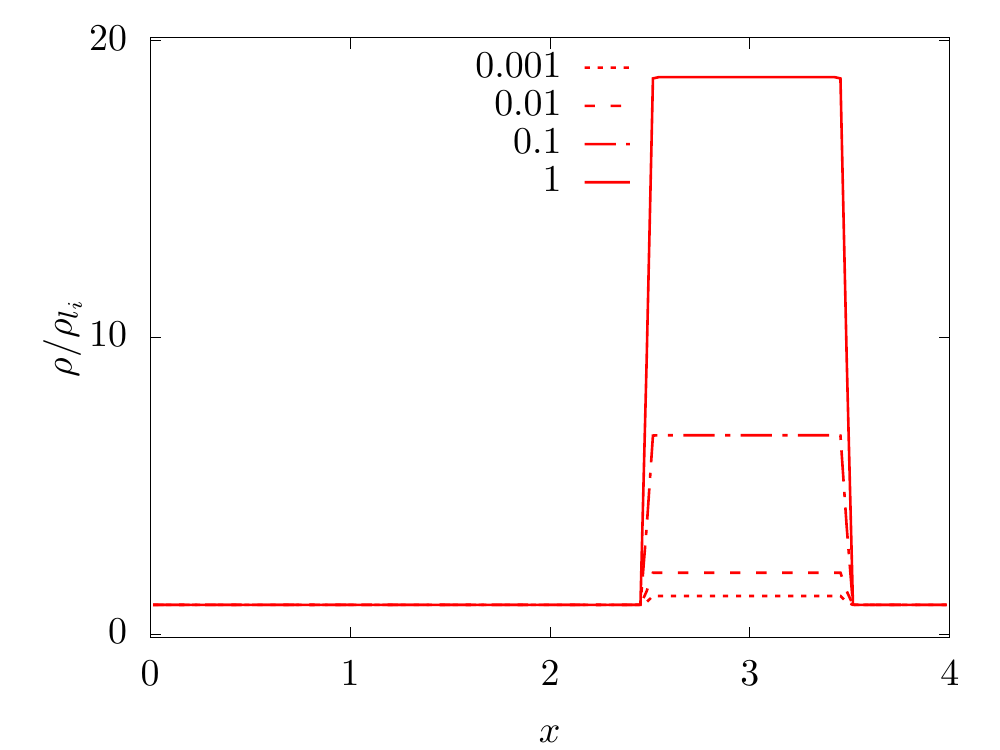}\label{fig_c1_c}}
\subfigure[]{\includegraphics[width=0.48\textwidth]{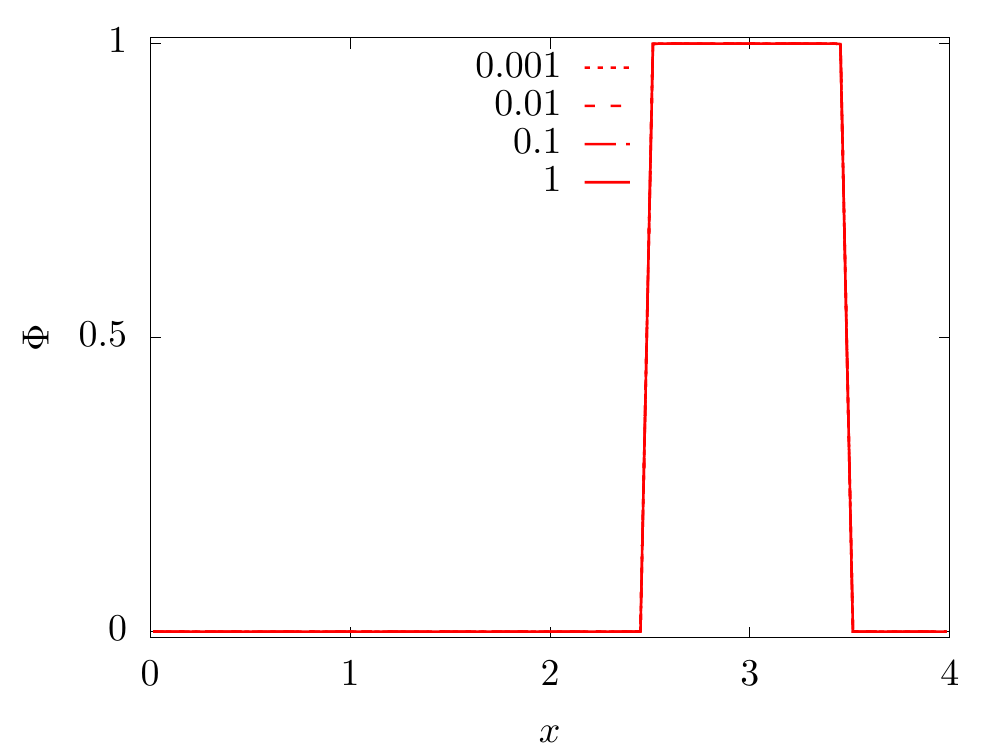}\label{fig_c1_d}}
\caption{a) Schematic of the computational domain and initial conditions for the test case \textit{C1a}. b) Non-dimensional thermodynamic pressure, $p_0/{\tilde p}$, in the gas regions as a function of the non-dimensional time, $t/{\tilde t}$. c)  Non-dimensional temperature, $T/T_{l,i}$, computed on the middle-line of the domain as a function of non-dimensional coordinate $x/{\tilde L}$.  d)  Non-dimensional  density, $\rho/\rho_{l,i}$, computed on the middle-line as a function of $x/{\tilde L}$. e) Volume fraction, $\Phi$, computed on the middle-line as a function of $x/{\tilde L}$. The temperature, density and volume fraction curves are provided for four different time instants, $t/{\tilde t}=0.001$, $t/{\tilde t}=0.01$, $t/{\tilde t}=0.1$ and $t/{\tilde t}=1$, ${\tilde t}$ being the reference time scale.}
\label{fig_c2}
\end{figure}
% FFFFFFFFFFFFFFFFFFFFFFFFFFFFFFFFFFFFFF
%
The geometrical configuration of the problem is provided in Fig.~\ref{fig_dmn_c1}. The ratio of the gas  to the liquid temperature is initially fixed to $(T_g/T_l)_i=5/6$. The initial temperature and density fields are uniform over each band, the only discontinuities being located on the liquid-gas interface. Fig.~\ref{fig_c2} provides also the temporal evolution of the thermodynamic pressure for the gas phase together with a plot of the density, temperature and volume fraction as a function of the axial position, $x/{\tilde L}$, at four different time instants. As a result of the initial temperature gradient, a heat flux develops from the liquid region towards the gas band. The temperature of the latter increases as shown in Fig.~\ref{fig_c1_a}, while its density decreases as can be observed in Fig.~\ref{fig_c1_b}. It should be noted that, the prescribed boundary conditions do not allow any volume change of the gas region. Hence, the transformation is isochoric, the volume fraction field remains unchanged and the gas band does not change the position of its centroid neither its boundaries during the transient as can be seen in Fig.~\ref{fig_c1_c}. In these conditions, the energy transfer to the gas band enforces the thermodynamic pressure to progressively increase until a uniform temperature field is established over the entire domain. At this point, the thermodynamic pressure settles to a constant value.\par
%
% FFFFFFFFFFFFFFFFFFFFFFFFFFFFFFFFFFFFFF
\begin{figure}[ht!]
\centering
\subfigure[]{\includegraphics[width=0.75\textwidth]{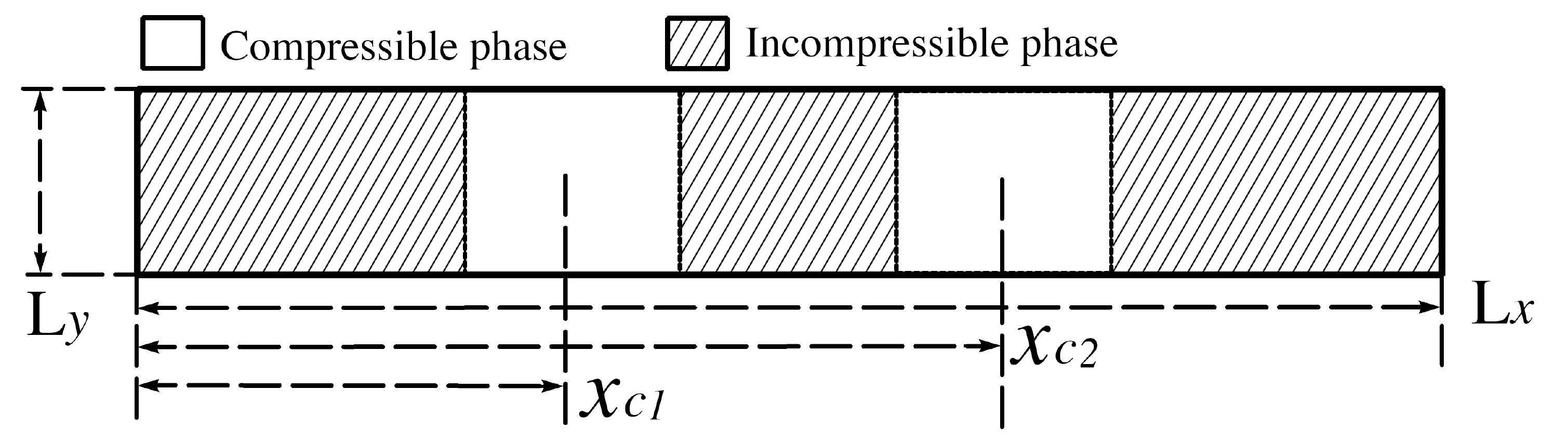}\label{fig_dmn_c2}}
\subfigure[]{\includegraphics[width=0.48\textwidth]{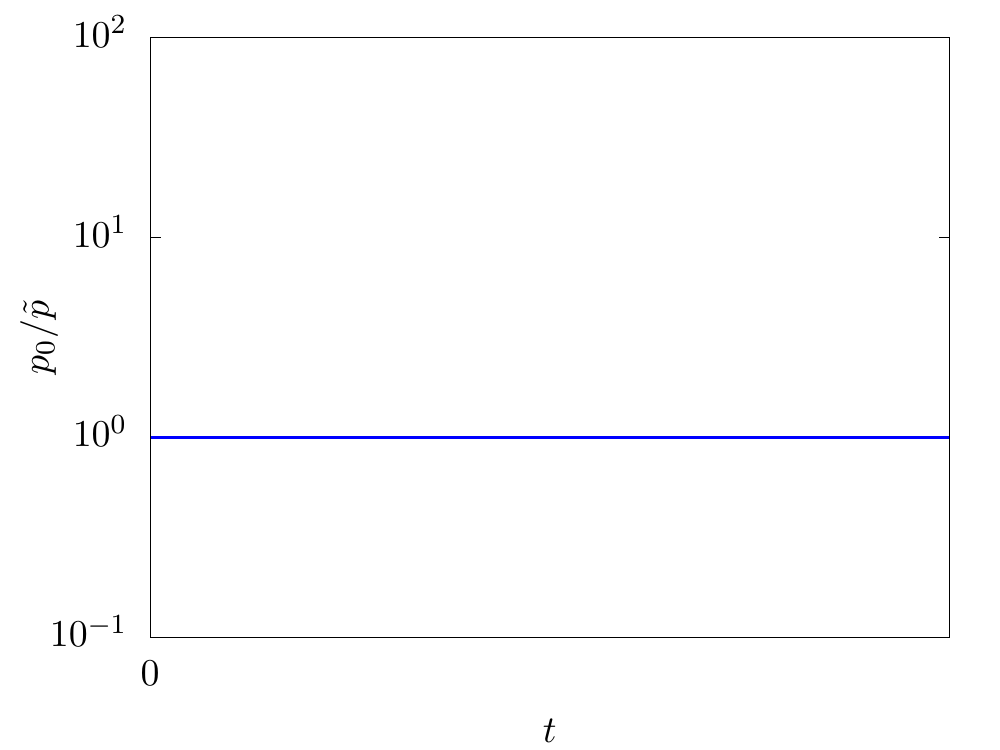}\label{fig_c2_a}}
\subfigure[]{\includegraphics[width=0.48\textwidth]{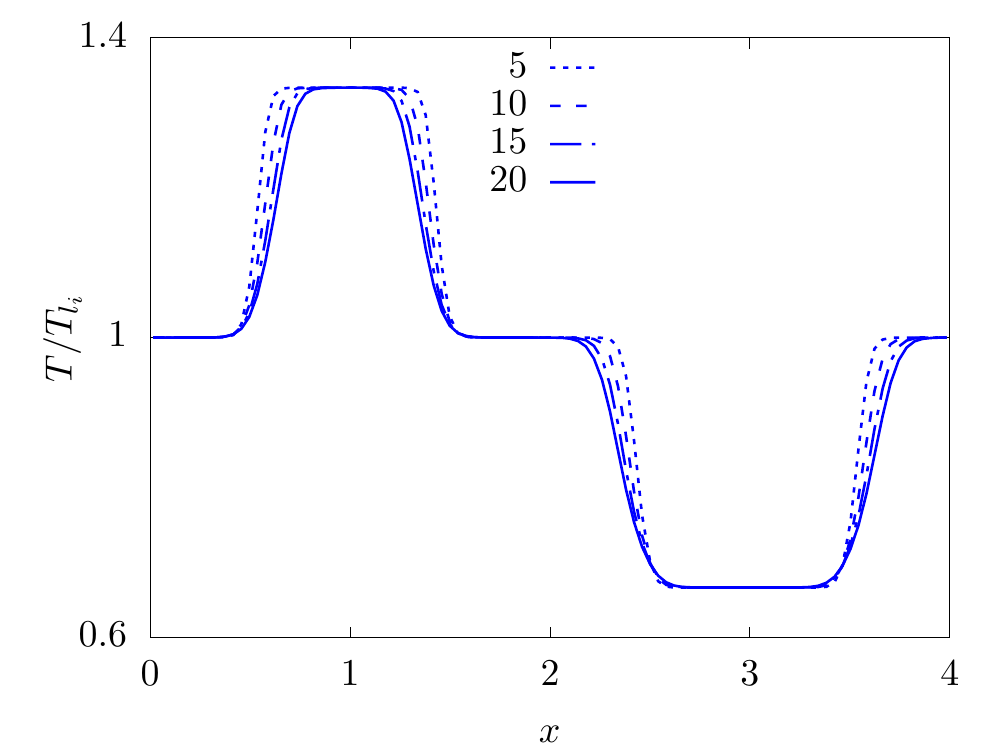}\label{fig_c2_b}}
\subfigure[]{\includegraphics[width=0.48\textwidth]{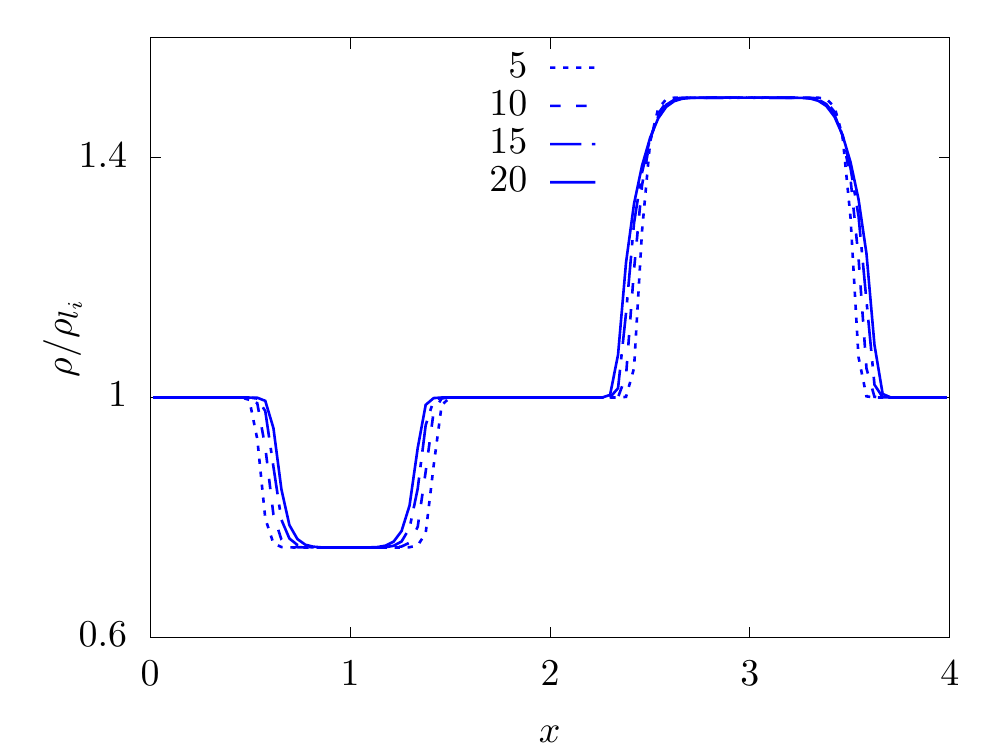}\label{fig_c2_c}}
\subfigure[]{\includegraphics[width=0.48\textwidth]{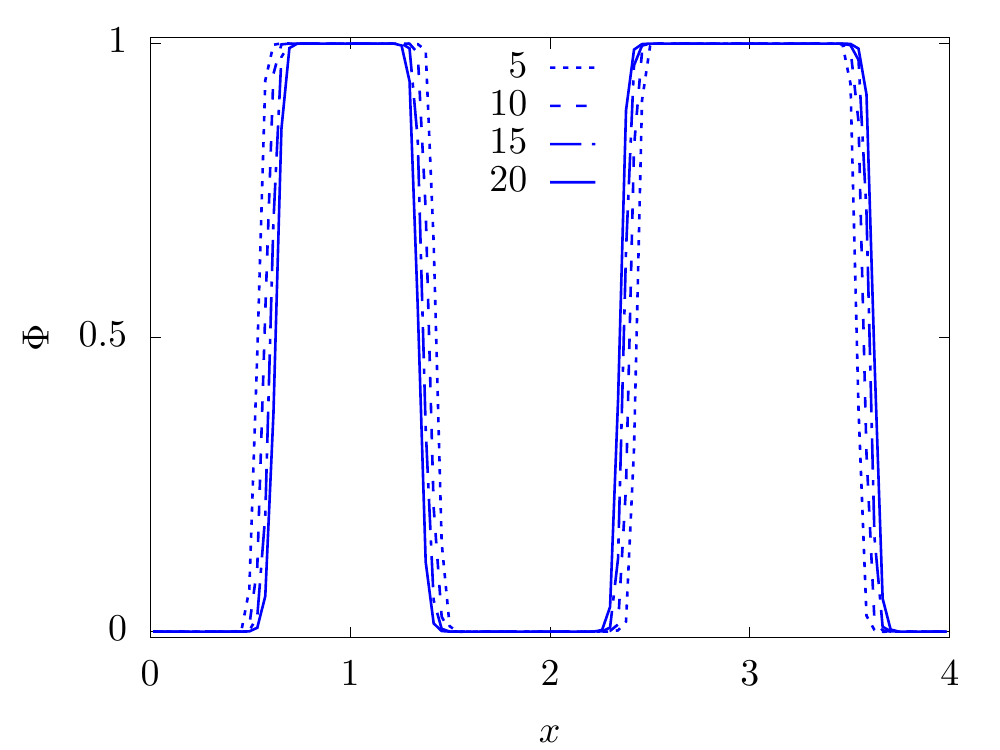}\label{fig_c2_d}}
\caption{a) Schematic of the computational domain and initial conditions for the test case \textit{C1b}. b) Thermodynamic pressure, $p_0/{\tilde p}$, in the gas regions as a function of the non-dimensional time, $t/{\tilde t}$. c) Temperature, $T/T_{l,i}$, as a function of $x/{\tilde L}$.  d)  Density, $\rho/\rho_{l,i}$, as a function of $x/{\tilde L}$. e) Volume fraction, $\Phi$, as a function of $x/{\tilde L}$. The temperature, density  and volume fraction are provided for four different time instants, $t/{\tilde t}=5$, $t/{\tilde t}=10$, $t/{\tilde t}=15$ and $t/{\tilde t}=20$.}
\label{fig_c3}
\end{figure}
% FFFFFFFFFFFFFFFFFFFFFFFFFFFFFFFFFFFFFFF
%
In the second test case, \textit{C1b}, we address the simulation of  the isobaric contraction and expansion of two separated gaseous bands enclosed within an incompressible liquid medium. The domain size, discretization and boundary conditions are unchanged and similarly for the fluid parameters which are provided in table~\ref{tab_param}. The initial configuration of the fluid system is provided in Fig.~\ref{fig_dmn_c2}. Initially, the two rectangular gas bands extend over a length  $b/{\tilde L}=1$ along the $x$ direction, separated by an incompressible liquid. The left-side band centroid is located at the axial position $x_{c1}/{\tilde L}=1$ while the right-side band is centered around $x_{c2}/{\tilde L}=3$. The ratio between the initial gas and liquid temperatures is set to $(T_{g,1}/T_l)_i=4/3$ for the left-side band and to $(T_{g,2}/T_l)_i=2/3$ for the right-side region. The initial temperature and density fields are uniform over each of the five different regions composing the fluid system. The initial thermodynamic pressure is the same in the two gaseous regions. Fig.~\ref{fig_c2_b}~and~\ref{fig_c2_c} provide the temperature and the density fields as a function of $x/{\tilde L}$ at four different time instants. The colder, right-side, band absorbs energy from the surrounding liquid medium while the hotter band on the left-side releases energy to the surrounding fluid. Hence, we observe the expansion of the colder gas band, simultaneously with the equivalent compression of the hotter gas, as shown in Fig.~\ref{fig_c2_d} providing the volume fraction as a function of $x/{\tilde L}$ at four different time instants. The volume of the liquid region included between the two gaseous bands cannot change; however, due to the periodic boundary condition along the $x$ direction, the liquid fluid can move from the right to left side of the domain. The two bands do not change the position of their center of mass during the expansion and contraction. Since in this case the volume of each band can vary freely, the transformation is isobaric as can be seen in Fig.~\ref{fig_c2_a}. Even if we cannot provide an analytical solution for the cases \textit{C1a} and \textit{C1b}, we believe that their numerical outcomes clearly show the capability of the method to account for heat transfer, density and temperature gradients as well as for compressibility effects in both isobaric and isochoric conditions in the low-Mach number regime.
\FloatBarrier
\subsection{Rising bubble}\label{sec:rising_bubble}\label{sec:case2}
\begin{figure}[hb!]
\centering
\includegraphics[width=0.6\textwidth]{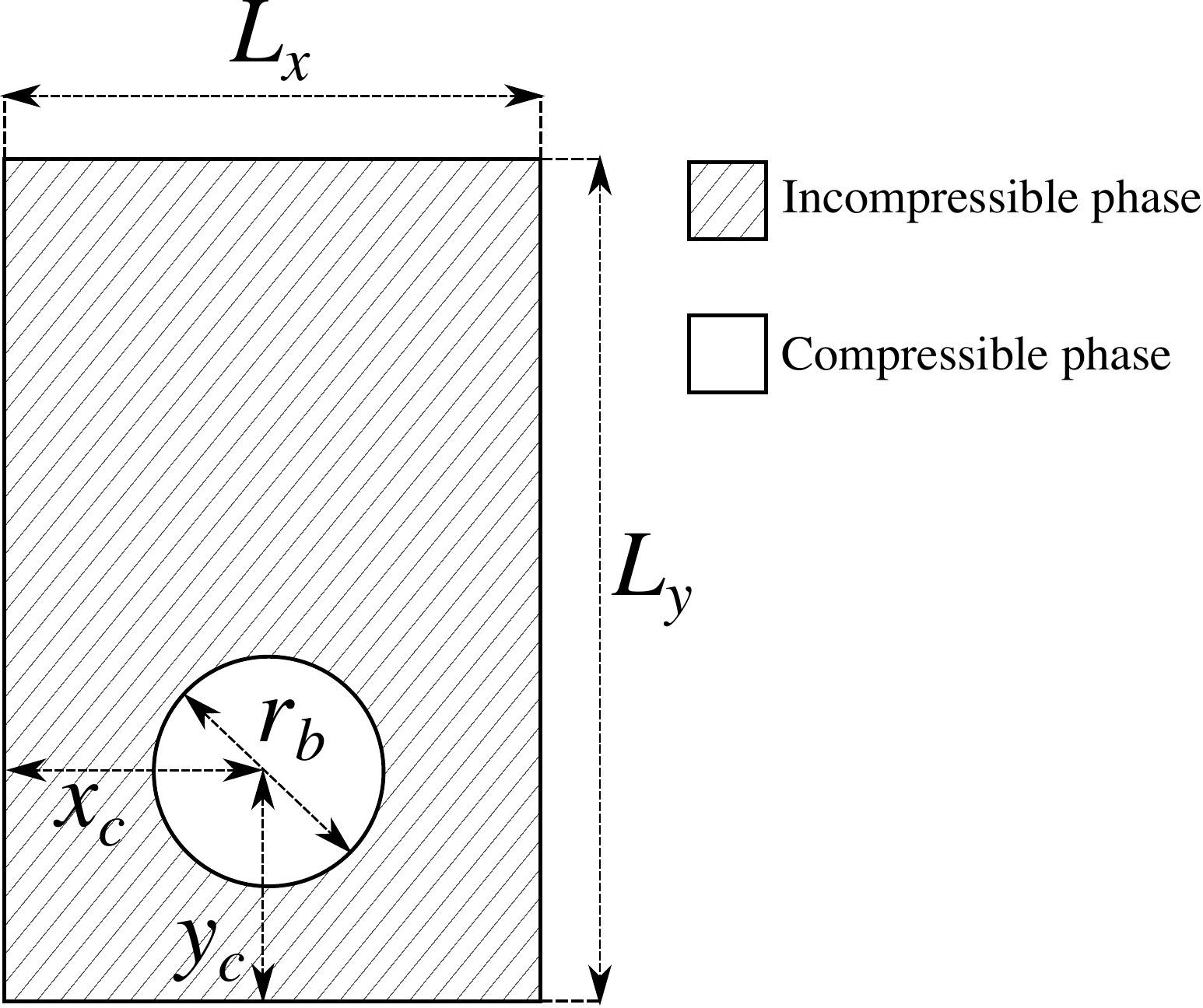}
\caption{Schematic of the computational domain and initial configuration used for the rising bubble simulation.}
\label{fig_dmn_c3}
\end{figure}

The test case \textit{C2} addresses the simulation of a two-dimensional rising bubble flow. A circular gaseous bubble of initial density $\rho_g$ and temperature $T_g$ is immersed in a liquid fluid with a higher, constant density, $\rho_l$, and temperature, $T_l$. Both the temperature and the density fields are initially uniform within the bubble and the liquid phase while a discontinuity exists across the interface. The initial configuration is displayed in Fig.~\ref{fig_dmn_c3}. 
%
% FFFFFFFFFFFFFFFFFFFFFFFFFFFFFFFFFFFFFF
% FFFFFFFFFFFFFFFFFFFFFFFFFFFFFFFFFFFFFF
%
The rectangular computational domain extends, in non-dimensional units, for $L_x/{\tilde L}\times L_y/{\tilde L}=1\times 2$. The domain is discretized using $N_x\times N_y=128 \times 256$ nodes. The initial bubble diameter is $d_{i}/{\tilde L}=0.5$ while the bubble center is initially located at $\mathbf{X}_{c,i}/\tilde{L} = (0.5,0.5)$. A no-slip and no-penetration boundary condition is prescribed to the momentum equation along the lower and the upper edges of the domain while a zero-gradient boundary condition is applied to the temperature equation. A periodic boundary condition is prescribed along the $x$ direction. The physical parameters of the fluids are provided in table~\ref{tab_param}. We report the results of five different test cases with different initial temperature ratios, $(T_l/T_g)_i= 1$, $1.2$, $1.5$, $2$ and $3$, and corresponding density ratios, $(\rho_l/\rho_g)_i = 10$, $8.33$, $6.67$, $5$ and $3.33$. We use as output quantities the center of mass of the bubble and the bubble rising velocity. The bubble centroid is defined as
%
% EEEEEEEEEEEEEEEEEEEEEEEEEEEEEEEEEEE
\begin{equation}
\mathbf{X}_c=(x_c,y_c)=\frac{\int_{V_g} \mathbf{x} dV}{\int_{V_g} dV}.
\end{equation}
% EEEEEEEEEEEEEEEEEEEEEEEEEEEEEEEEEEE
%
In a similar fashion, the bubble rising velocity is defined as the mean velocity with which the gas phase is moving,
%
% EEEEEEEEEEEEEEEEEEEEEEEEEEEEEEEEEEE
\begin{equation}
\mathbf{U}_c=(u_c,v_c)=\frac{\int_{V_g} \mathbf{u} dV}{\int_{V_g} dV}.
\label{rising_vel}
\end{equation}
% EEEEEEEEEEEEEEEEEEEEEEEEEEEEEEEEEEE
%
% FFFFFFFFFFFFFFFFFFFFFFFFFFFFFFFFFFFFFF
\begin{figure}[ht!]
\centering
\includegraphics[width=0.8\textwidth]{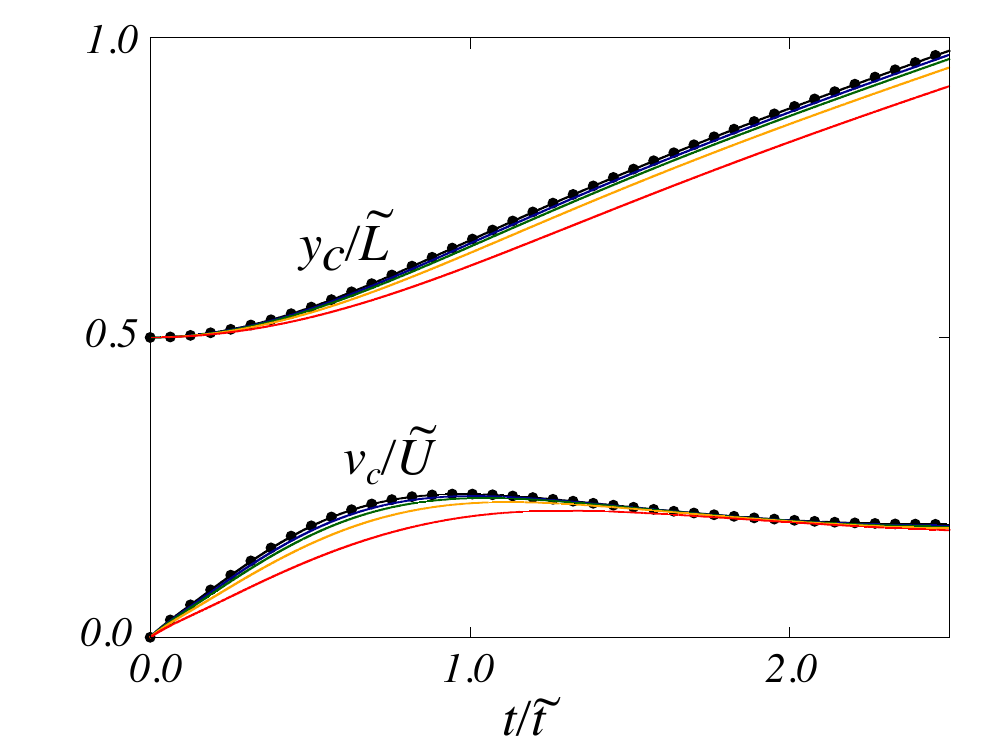}
\caption{Rising bubble: vertical position of the bubble centroid, $y_c(t)/{\tilde L}$, and vertical component of the bubble rising velocity, $v_c(t)/{\tilde U}$, versus time, $t/{\tilde t}$. The continuous black line represents the outcomes of the present isothermal case whereas the dotted line provides the results by Hysing et al.~\cite{hysing2009quantitative}.}
\label{fig_c1}
\end{figure}
% FFFFFFFFFFFFFFFFFFFFFFFFFFFFFFFFFFFFFF
%
Fig.~\ref{fig_c1} displays the vertical position of the bubble centroid, $y_c(t)/{\tilde L}$, and the vertical component of the bubble rising velocity, $v_c(t)/{\tilde U}$, versus time, $t/{\tilde t}$, for each of the initial temperature ratios given above. In the isothermal case the initial temperature field is uniform over the entire domain, $(T_l/T_g)_i=1$, and the density ratio is set to $(\rho_l/\rho_g)_i=10$. The results of the present simulation are compared with that obtained by Hysing et al.~\cite{hysing2009quantitative}. As the density ratio, $(\rho_l/\rho_g)_i$, is decreased, the  rising velocity of the bubble is initially lower than in the isothermal case due to the lower buoyancy force exerted by the liquid on the gas bubble. Nonetheless, the thermal diffusion reduces progressively the temperature gradient between the two phases. The bubble heats-up and the density ratio, $\rho_l/\rho_g$, increases. As a result, after an initial transient, the terminal bubble rising velocity tends to settle to the same regime velocity as that of the isothermal case, independently of the initial density ratio, $(\rho_l/\rho_g)_i$. Clearly, the initial differences in the rising velocities lead to an offset in the position of the bubble centroid.\par
\begin{figure}[t!]
  \centering
  \includegraphics[width=0.5\textwidth]{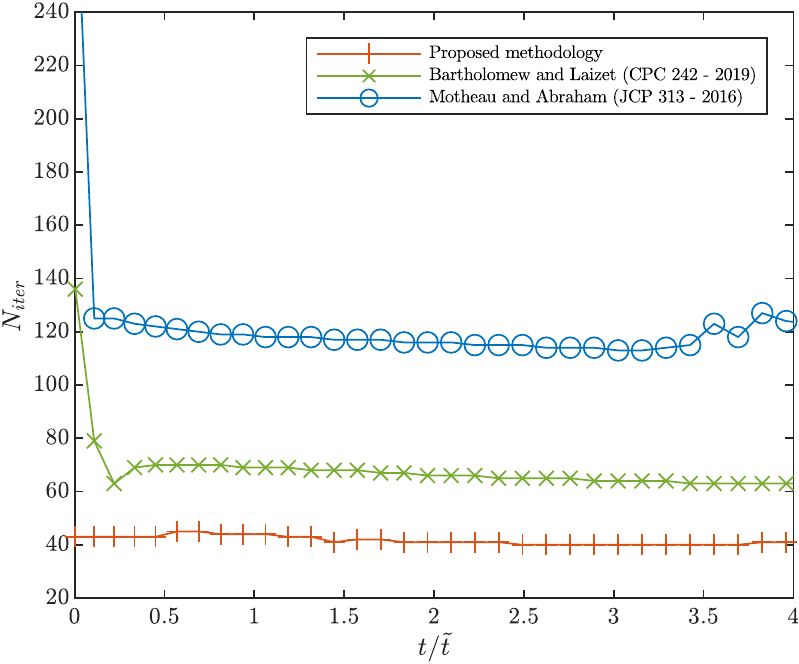}
  \caption{\scapin{Number of iterations as a function of time $t/\tilde{t}$ required to solve the pressure Poisson equation for the rising bubble test case using the method by Bartholomew and Laizet~\cite{bartholomew2019new}, Motheau and Abraham~\cite{motheau2016high} and the present method. The reference time scale $\tilde{t}$ is set equal to $\sqrt{d_0/\mathbf{g}}$.}}
  \label{fig_comp}
\end{figure}

To conclude the analysis of this test case, we compare the three methods presented in section~\ref{sec:flow_solver} in terms of number of iterations needed to solve the Poisson equation. Since the three methods require different tolerances to satisfy the divergence constraint with the same accuracy, we set $\varepsilon_t=10^{-8}$ for \textit{Method 1} and $\varepsilon=10^{-11}$ for \textit{Method 2 and 3} for a fair comparison. Using these different thresholds, $\varepsilon_t$, leads to similar values of the residual (below $10^{-14}$), computed as the difference between the velocity divergence and its constraint according to Eq.~\eqref{eq_34}. For what concerns \textit{Method 3}, and only for this case, we compute the residual $\varepsilon$ on the pressure between two consecutive iterations and not on the velocity divergence. Once again, this choice is motivated by a fair comparison with the other two methods. Indeed, for \textit{Method 1} and \textit{Method 2}, the residual based on the pressure $p_2$ is the only possible choice being the correction step performed only at the end. The results provided in Fig.~\ref{fig_comp} clearly show that the current approach, \textit{Method 3}, requires a number of iterations between $1.5$ and $3$ times lower than that of \textit{Method 1} and \textit{Method 2} to achieve a full convergence. As mentioned above, we attribute this faster convergence to the exact way we recast the variable-coefficient Poisson equation into a constant-coefficient problem using directly the correction step.
%
% FFFFFFFFFFFFFFFFFFFFFFFFFFFFFFFFFFFFFF

% FFFFFFFFFFFFFFFFFFFFFFFFFFFFFFFFFFFFFF
%
\subsection{Multiple rising bubbles}\label{sec:case3}
\begin{figure}[t!]
\centering
\includegraphics[width=0.7\textwidth]{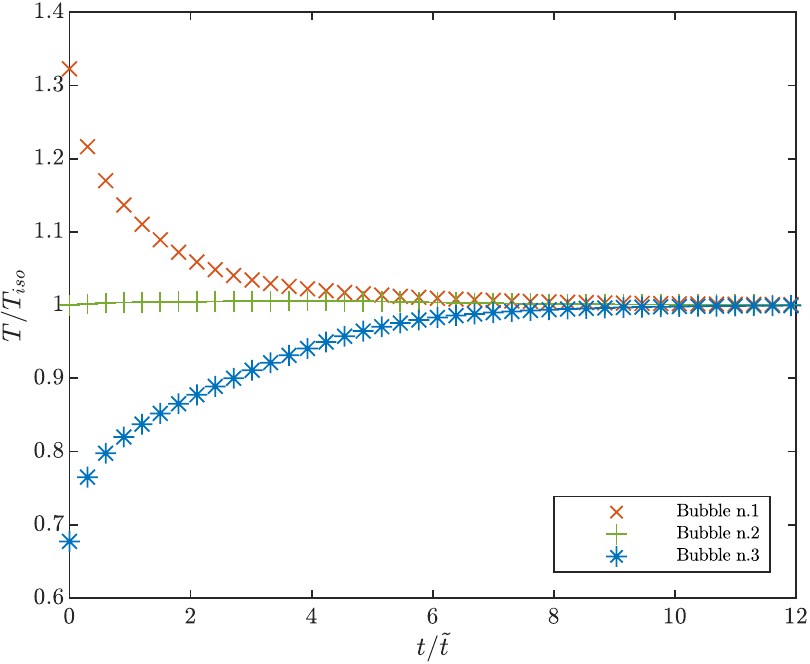}
\caption{\scapin{Position of the three rising bubbles at $t/\tilde{t}=0$ (left panel) and for $t/\tilde{t}>0$ (right panel). The interface position is taken from the grid points where $\Phi=0.5$ and the bubble contour are plotted with at the dimensionless physical time $t/\tilde{t}=\{0,\,1.5,\,3.0,\,4.5,\,6.0,\,7.5,\,9.0,\,10.5,\,12.0,\,13.5\}$, with the reference time scale being $\tilde{t}=\sqrt{d_0/|\mathbf{g}|}$.}}
\label{fig_multiple_bub}
\end{figure}

In this section, we consider the same configuration adopted in the previous test case to study the flow in the presence of three compressible bubbles, rising in an incompressible liquid. The bubbles have the same initial diameter, $d_i$ and are initially at rest in a rectangular domain of dimensions $L_x/\tilde{L}\times L_y/\tilde{L}=7.2\times12.8$, being the reference length $\tilde{L}=d_i$. The initial position of the bubble centroids are set to $(\mathbf{X}_{c,1}/\tilde{L})_i = (1.8, 1.25)$ for \textit{Bubble n.1}, $(\mathbf{X}_{c,2}/\tilde{L})_i = (3.6, 1.25)$ for \textit{Bubble n.2} and $(\mathbf{X}_{c,3}/\tilde{L})_i  = (5.4, 1.25)$ for \textit{Bubble n.3}. The initial temperature and density of the liquid phase are $T_l$ and $\rho_l$, respectively. In order to highlight the compressibility effects, the three bubbles are initialized at three different temperatures, $(T_{g,1}/T_l)_i=1.5$, $(T_{g,2}/T_l)_i=1.0$ and $(T_{g,3}/T_l)_i=0.75$ as reported on the left panel of Fig.~\ref{fig_multiple_bub}. In order to avoid bubble coalescence and merging, we consider a limited Weber number $We=\rho_{g,0}^m\tilde{U}^2d_0/\tilde{\sigma}=0.125$ and we set the Reynolds number $Re=\rho_{g,0}^m\tilde{U}d_0/\mu_g=125$, where $\tilde{U}=\sqrt{|\mathbf{g}|d_0}$ and $\rho_{g,0}^m$ is the minimum initial gas density. The Prandtl number $Pr=\mu_gk_g/c_{p,g}$ is set to $0.7$. All the other dimensionless parameters are kept the same as in the previous case and are reported in table~\ref{tab_param}.
%
% FFFFFFFFFFFFFFFFFFFFFFFFFFFFFFFFFFFFFF
% FFFFFFFFFFFFFFFFFFFFFFFFFFFFFFFFFFFFFF
%
% FFFFFFFFFFFFFFFFFFFFFFFFFFFFFFFFFFFFFF
\begin{figure}[t!]
\centering
\includegraphics[width=0.45\textwidth, height=5 cm]{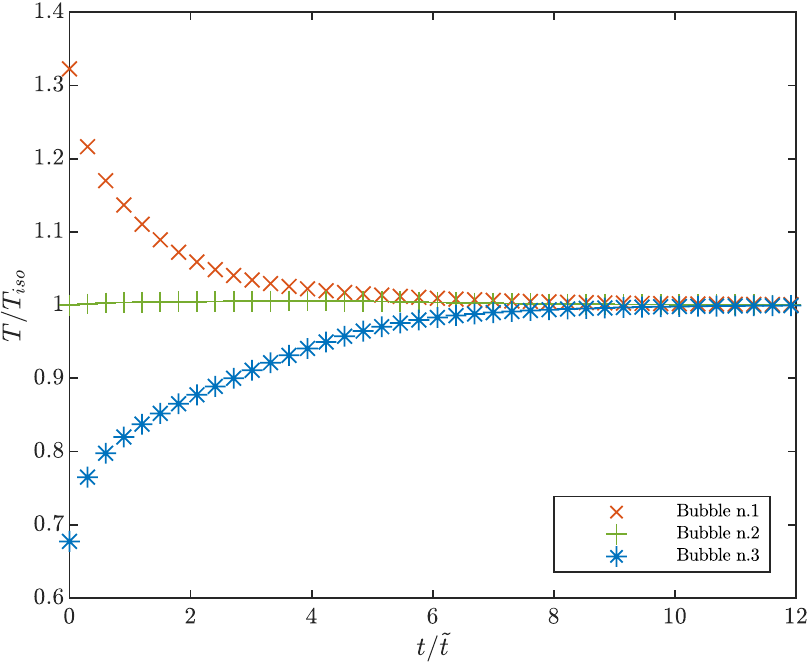}
\includegraphics[width=0.45\textwidth, height=5 cm]{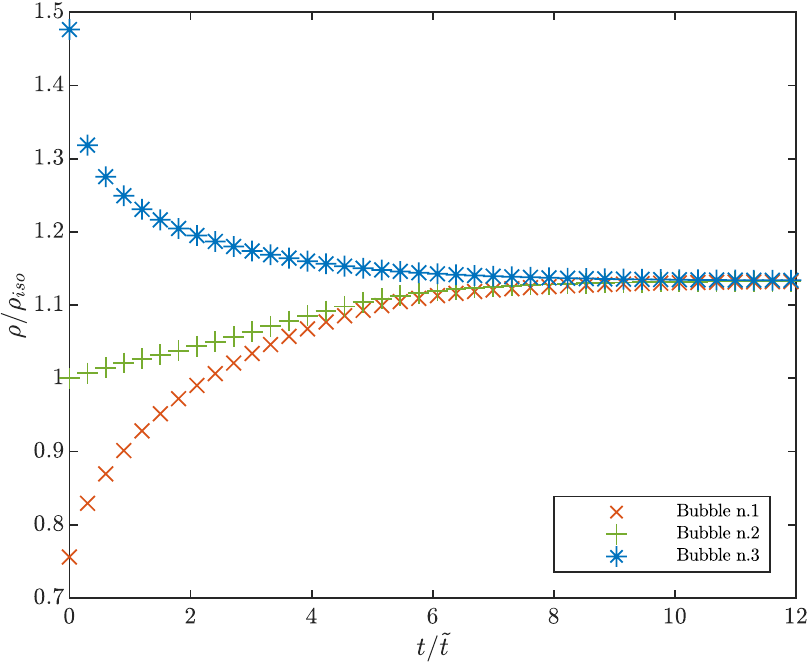}
\caption{Averaged gas temperature (left panel) and averaged gas density of the three bubbles (right panel) versus non-dimensional time. The initial temperature and density of \textit{Bubble n.2} are taken as a reference temperature, $T_{iso}$, and density, $\rho_{iso}$.  The reference time scale is $\tilde{t}=\sqrt{d_0/|\mathbf{g}|}$.}
\label{fig_tmp_rho}
\end{figure}
% FFFFFFFFFFFFFFFFFFFFFFFFFFFFFFFFFFFFFF
%
Since the system is closed and thermally isolated, once the bubbles start to rise, the heat transfer exchanged among each other and with the liquid medium drives them towards the thermal equilibrium. In detail, \textit{Bubble n.1} starts to cool down, \textit{Bubble n.2} maintains an almost constant average temperature, whereas \textit{Bubble n.3} is heated up. As a result and owing to the variation of the thermodynamic pressure, the first bubble contracts increasing its mean density, the third bubble expands decreasing its mean density, whereas the second one slightly expands mainly due to the variation of the thermodynamic pressure. As shown in Fig.~\ref{fig_tmp_rho}, after $t/\tilde{t}\approx 6$, being $\tilde{t}=\sqrt{d_0/|\mathbf{g}|}$, the thermal equilibrium is globally reached and the mean temperature and density remain approximately constant for the three bubbles.\par
%
% FFFFFFFFFFFFFFFFFFFFFFFFFFFFFFFFFFFFFF
\begin{figure}[t!]
\centering
\includegraphics[width=0.60\textwidth]{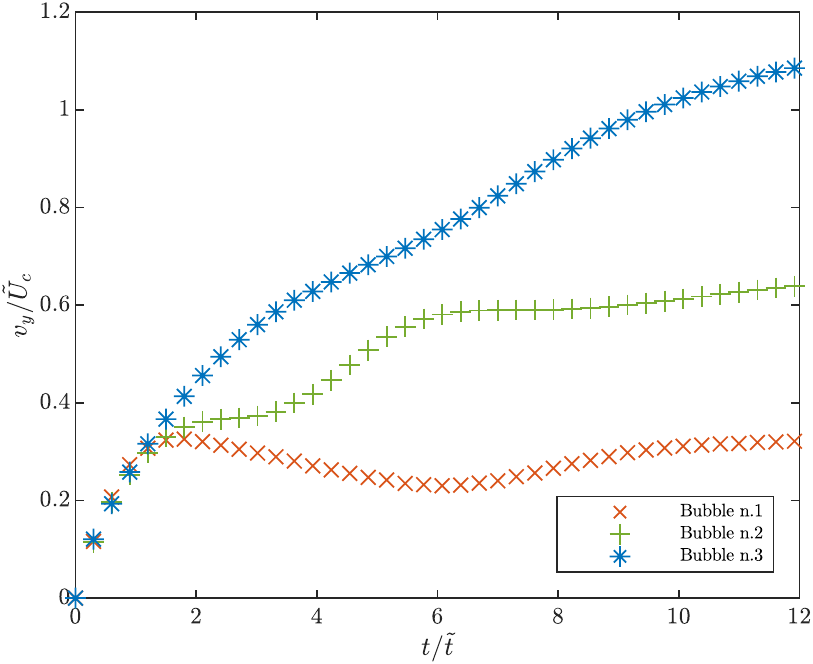}
\caption{Normalized vertical velocity of the bubbles versus time. The reference velocity scale is $\tilde{U}=\sqrt{|\mathbf{g}|d_0}$ while the reference time scale is $\tilde{t}=\sqrt{d_0/|\mathbf{g}|}$.}
\label{fig_rising_vel}
\end{figure}
% FFFFFFFFFFFFFFFFFFFFFFFFFFFFFFFFFFFFFF
%
Fig.~\ref{fig_rising_vel} provides the mean vertical velocity of the centroid of each bubble. The initial expansion and contraction of the bubbles affects the vertical component of their rising velocities computed as in Eq.~\eqref{rising_vel}. In particular, until $t/\tilde{t}\approx 1.8$, all the three bubbles move with a comparable vertical velocity. After this initial stage, the third bubble starts to accelerate and arrives first at the top wall, whereas the first one starts to decelerate and moves along the vertical direction at an almost constant speed. On the other hand, the second bubble accelerates towards the top wall, but at a lower rate than the third one. The physical explanation for this behavior relies in the modification induced by the initial expansion and contraction stage of the bubbles, which determines an increase of the buoyancy forces for \textit{Bubble n.3} and a reduction for \textit{Bubble n.1}.
\subsection{Mixing layer}\label{sec:case4}
As a final test case, \textit{C4}, the numerical simulation of a two-dimensional, temporal mixing layer is addressed. This considered flow configuration develops in the region between two counter-directional flows, one of them being compressible, the other incompressible. The streams move with opposite velocities, $U_g$ and $U_l$. In these conditions, a Kelvin-Helmholtz instability promotes the formation of well-defined coherent vortices in the region separating the two streams. The latter enhance micro-mixing and molecular diffusion promoting the exchange of momentum and energy between the opposite streams. The computational domain consists of a square box of unit size, $L_x/\tilde{L}\times L_y/\tilde{L} = 1\times 1$, discretized using $N_x \times N_y=512 \times 512$ nodes. In the lower part of the computational domain, $0<y/{\tilde L}\le0.5$, the incompressible flow moves from the right to the left while in the upper part of the domain, $0.5<y/{\tilde L}\le 1$, the compressible stream moves in the opposite direction. A no-slip boundary condition is prescribed to the momentum equation along the top and bottom sides of the domain while a zero-gradient, adiabatic boundary condition is assigned to the temperature equations along the same boundaries. Periodic boundary conditions are applied to all quantities along the flow direction, $x$. \par
To better characterize the mixing-layer flow, it is worth introducing a length-scale based on the initial vorticity thickness in the mixing layer, $\delta$, and a corresponding Reynolds number $Re_\delta=U_c \delta/\nu_{g,i}$ with $\nu_{g,i}$ the kinematic viscosity of the gas phase (evaluated at the initial condition) and $U_c$ a prescribed convective velocity defined as $U_c=1/2(U_g-U_l)$. The initial velocity field is prescribed imposing a pseudo-perturbation on a mean profile according to the following relations~\cite{zayernouri2011coherent}:
%
% EEEEEEEEEEEEEEEEEEEEEEEEEEEEEEEEEEE
\begin{align}
& \dfrac{u(x,y,0)}{U_c}=\tanh\left(\frac{2{\tilde L}}{\delta}y\right)+\xi_{noise}\frac{\partial\psi}{\partial y},\label{eq_10000}\\
& \dfrac{v(x,y,0)}{U_c}=-\xi_{noise}\frac{\partial\psi}{\partial x},\label{eq_10001}\\
&\psi(x,y)=\exp\left(-\frac{{\tilde L}^2}{\delta^2}y^2\right)\left[\cos(4\pi x)+0.03\sin(10\pi x)\right],
\end{align}
% EEEEEEEEEEEEEEEEEEEEEEEEEEEEEEEEEEE
%
where $u(x,y,0)$ and $v(x,y,0)$ are the horizontal and vertical components of the initial velocity field. Moreover, the factor $\xi_{noise}=10^{-3}$ is chosen to ensure that the velocity perturbations remain a small percentage of the mean velocity, as suggested by the authors in Zayernouri~et~al.~\cite{zayernouri2011coherent}. Prescribing the hyperbolic tangent velocity profile given by Eq.~\eqref{eq_10000}~-~\eqref{eq_10001}, the wave-length associated with the initial vortex distribution results to be approximately  $\lambda\simeq7\delta$~\cite{betchov1963stability,zayernouri2011coherent}. Hence, given the domain size, ${\tilde L}$ and the desired number of vortexes in the periodic domain, $N$, the initial vortex thickness is $\delta/{\tilde L}=1/(7N)$. In the present case, the initial vorticity thickness is fixed to $\delta/{\tilde L}=1/28$ and $Re_\delta=200$. The non-dimensional viscosity, thermal conductivity and the specific heat capacity ratios are kept equal to unity, while the density ratio based on the initial gas density $(\rho_l/\rho_{g})_i$ is taken equal to $5$. Finally, the Prandtl number is set to $Pr=\mu_gc_{p,g}/k_g=8.92$ with $c_{p,g}$ and $k_g$ being the specific heat capacity and thermal conductivity of the gas phase. The initial temperature field is initialised according to the step-function,
%
% EEEEEEEEEEEEEEEEEEEEEEEEEEEEEEEEEEE
\begin{equation}
T(x,y)=
\begin{cases}
T_{l,i}, & \mbox{if}\ 0 \le y/{\tilde L}\le 0.5,\\ 
T_{g,i}, & \mbox{if}\ 0.5<y/{\tilde L}\le 1.
\end{cases}
\end{equation}
% EEEEEEEEEEEEEEEEEEEEEEEEEEEEEEEEEEE
%
% FFFFFFFFFFFFFFFFFFFFFFFFFFFFFFFFFFFFFF
\begin{figure}[t]
\centering
\includegraphics[width=0.45\textwidth]{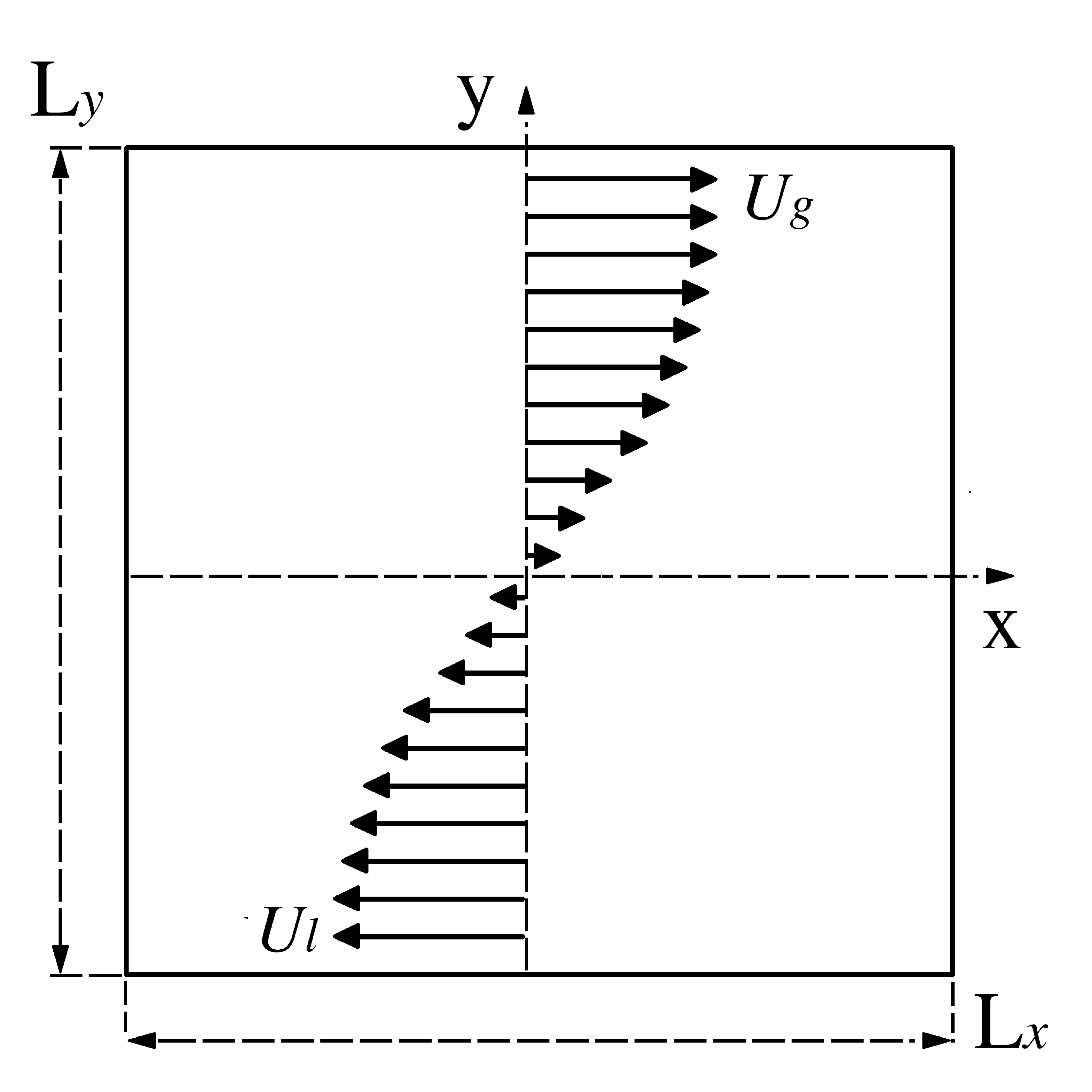}
\includegraphics[width=0.45\textwidth]{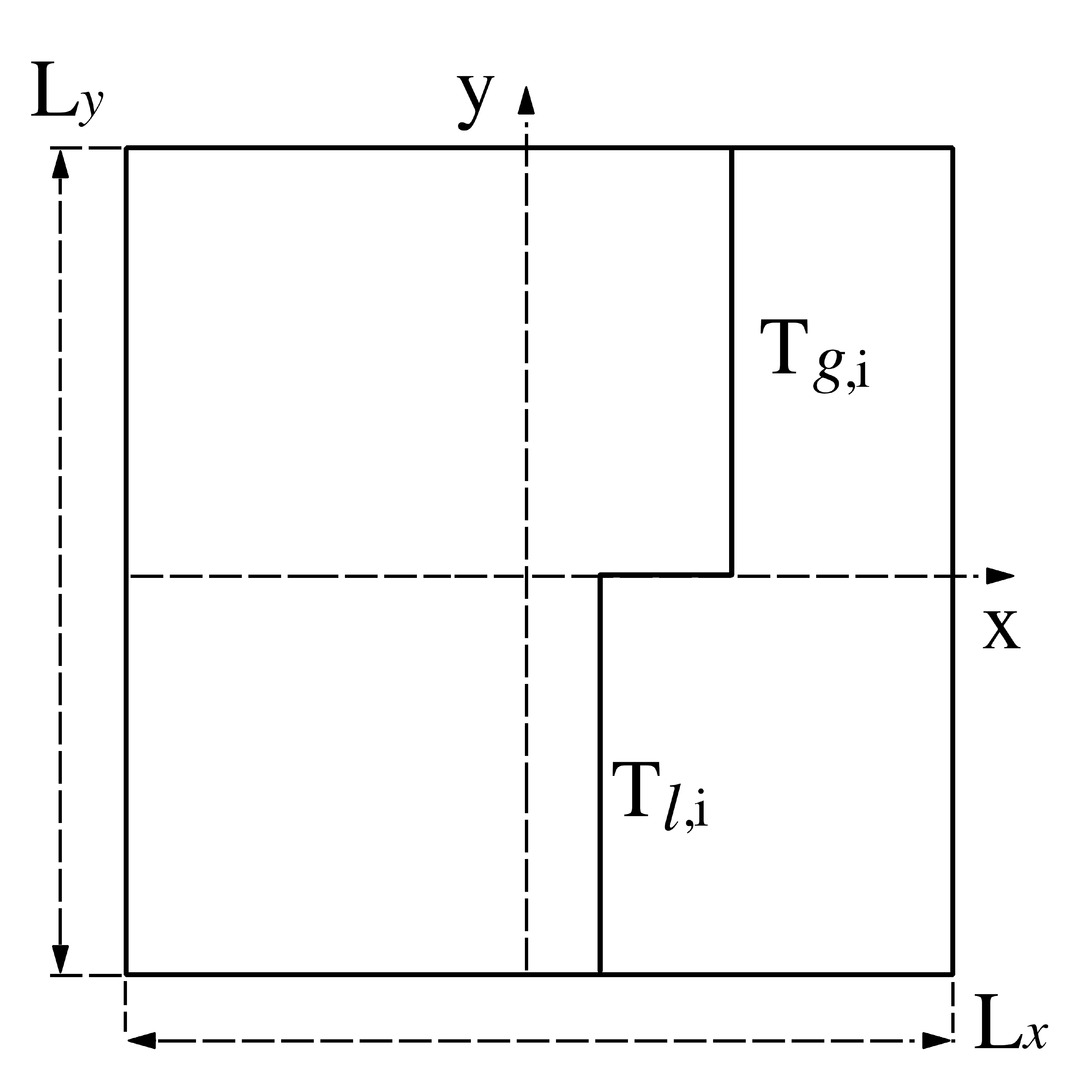}
\caption{Sketch of the domain for the temporal mixing-layer simulation showing the initial velocity and temperature fields.}
\label{fig_c4_1}
\end{figure}
% FFFFFFFFFFFFFFFFFFFFFFFFFFFFFFFFFFFFFF
%
While keeping fixed the initial liquid temperature $T_{l,i}$ and the initial density ratio $(\rho_l/\rho_{g})_i$ between the two phases, different initial gas temperatures, $T_{g,i}$, gas densities, $\rho_{g,i}$ and liquid densities $\rho_{l,i}$, are prescribed to the compressible and incompressible fluids, as sketched in Fig.~\ref{fig_c4_1}. A first test case considers the isothermal flow where $(T_g/T_l)_i=1$, whereas three other cases address a temperature ratio equal to $15/16$, $5/6$ and $3/4$, respectively. 
%
% FFFFFFFFFFFFFFFFFFFFFFFFFFFFFFFFFFFFFF
\begin{figure}[t]
\centering
\includegraphics[width=0.45\textwidth, height=5 cm]{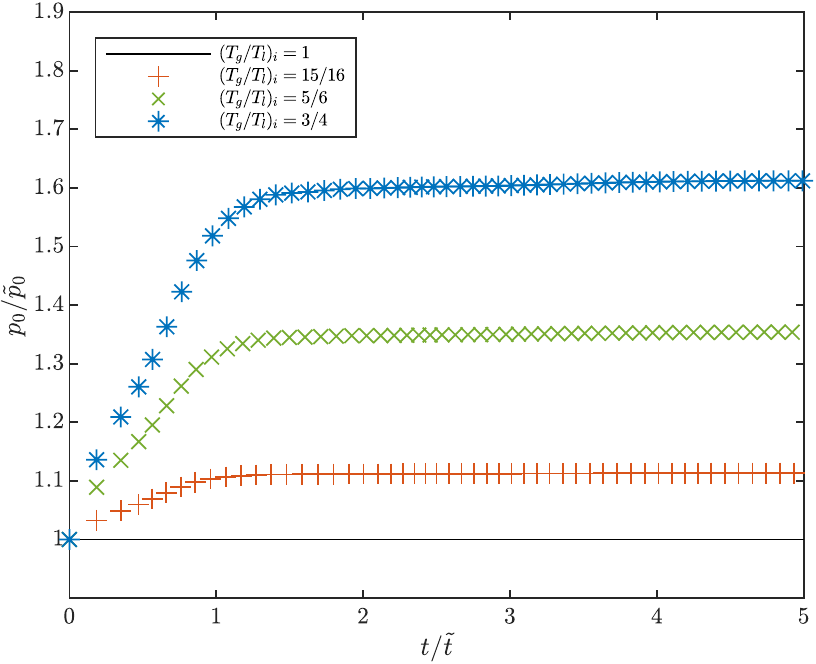}
\includegraphics[width=0.45\textwidth, height=5 cm]{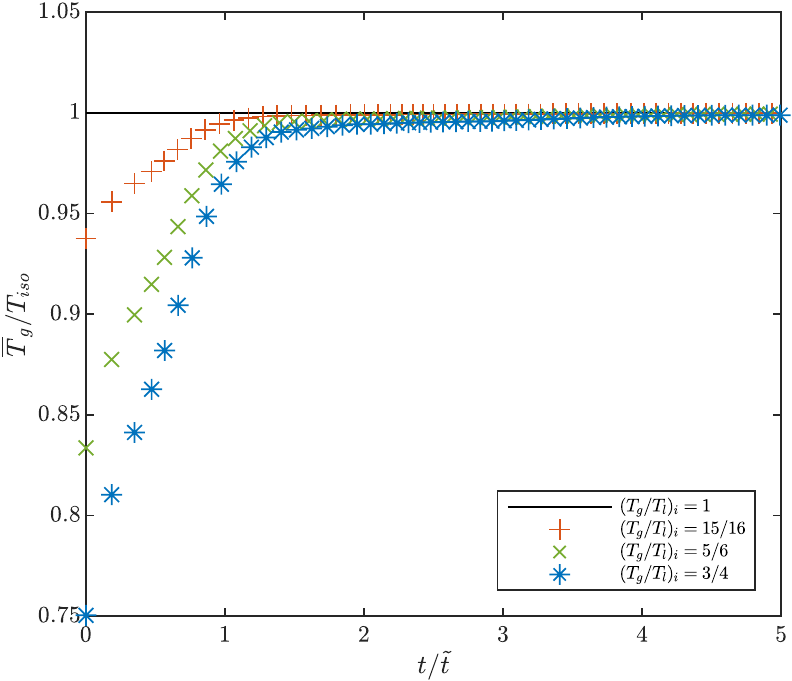}
\includegraphics[width=0.45\textwidth, height=5 cm]{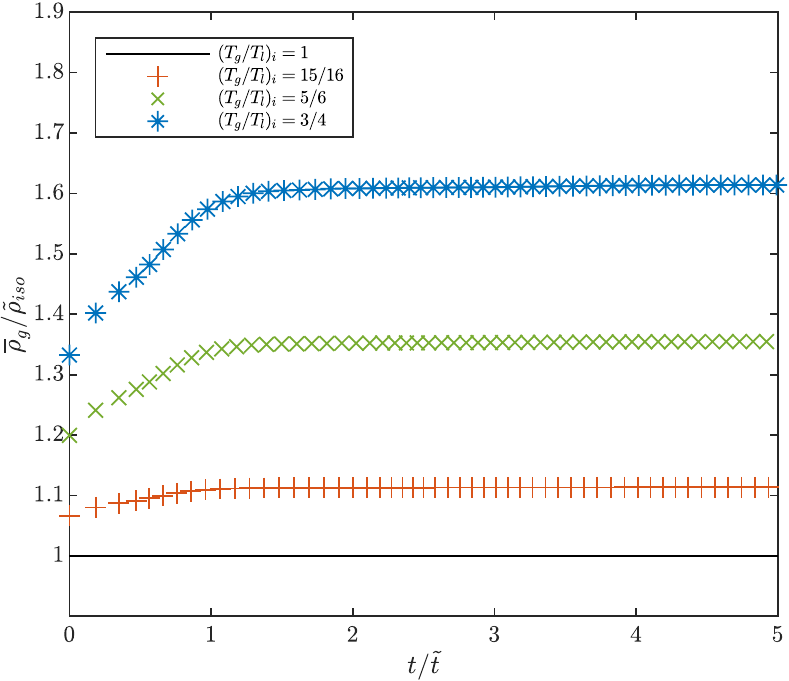}
\includegraphics[width=0.45\textwidth, height=5 cm]{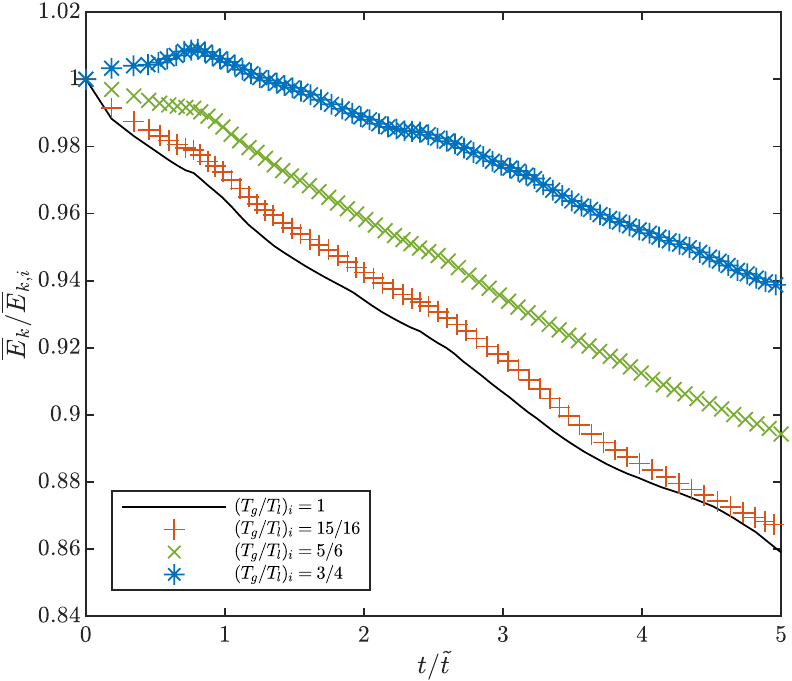}
\caption{\scapin{Temporal evolution of the thermodynamic pressure, $p_0$, mean gas temperature, $\overline{T}_g$, mean gas density, $\overline{\rho}_g$ and mean kinetic energy, $\overline{E}_k$ for four different temperature ratios, $(T_g/T_l)_i\in [1.0,15/16,5/6,3/4]$. All the quantities are non-dimensional using as reference values the quantities of the isothermal case, except for the mean kinetic energy where we employ the initial value of $\overline{E}_k$ of each case, $\overline{E}_{k,i}$. The reference time-scale $\tilde{t}$ is set equal to $\tilde{L}/\tilde{U}_c$}.}
\label{fig_c4_2}
\end{figure}
% FFFFFFFFFFFFFFFFFFFFFFFFFFFFFFFFFFFFFF
%
Fig.~\ref{fig_c4_2} provides the temporal evolution of the thermodynamic pressure (uniform over the computational domain), the mean gas and liquid temperature, the mean gas density and the mean kinetic energy for the two phases. The gas and liquid temperatures are computed as integrals over the corresponding domains, whereas the mean kinetic energy is estimated over the compressible and incompressible regions:
%
% EEEEEEEEEEEEEEEEEEEEEEEEEEEEEEEEEEE
\begin{align}
&{\overline{T}}_g(t)=\frac{1}{V_g}\int_{V_g} T(x,y,t) \Phi(x,y,t)\ dx\ dy,\\
&{\overline{T}}_l(t)=\frac{1}{V_l}\int_{V_l} T(x,y,t) (1-\Phi(x,y,t))\ dx\ dy,\\
&{\overline{E}_k}(t)=\frac{1}{2L_x L_y}\int_{-L_x/2}^{L_x/2}\int_{-L_y/2}^{L_y/2} \rho(x,y,t){\mathbf u(x,y,t)}\cdot{\mathbf u}(x,y,t) \ dx\ dy,
\end{align}
% EEEEEEEEEEEEEEEEEEEEEEEEEEEEEEEEEEE
%
where $V_g$ is computed using Eq.~\eqref{eqn:v_gas} and $V_l=L_xL_y-V_g$, being the domain closed. Once the mean gas temperature is known, the mean gas density is computed directly from the equation of state whereas the liquid density is constant and equal to $\rho_l$,
%
% EEEEEEEEEEEEEEEEEEEEEEEEEEEEEEEEEEE
\begin{equation}
{\bar\rho}_g(t)=\dfrac{p_{0}(t)}{\mathcal{R}\overline{T}_g(t)}\mathrm{,} \hspace{0.5 cm} \text{and} \hspace{0.5 cm} {\bar\rho}_l(t)=\rho_l\mathrm{.}
\end{equation}
% EEEEEEEEEEEEEEEEEEEEEEEEEEEEEEEEEEE
%
% FFFFFFFFFFFFFFFFFFFFFFFFFFFFFFFFFFFFFF
\begin{figure}[h!]
\centering
\includegraphics[width=0.40\textwidth]{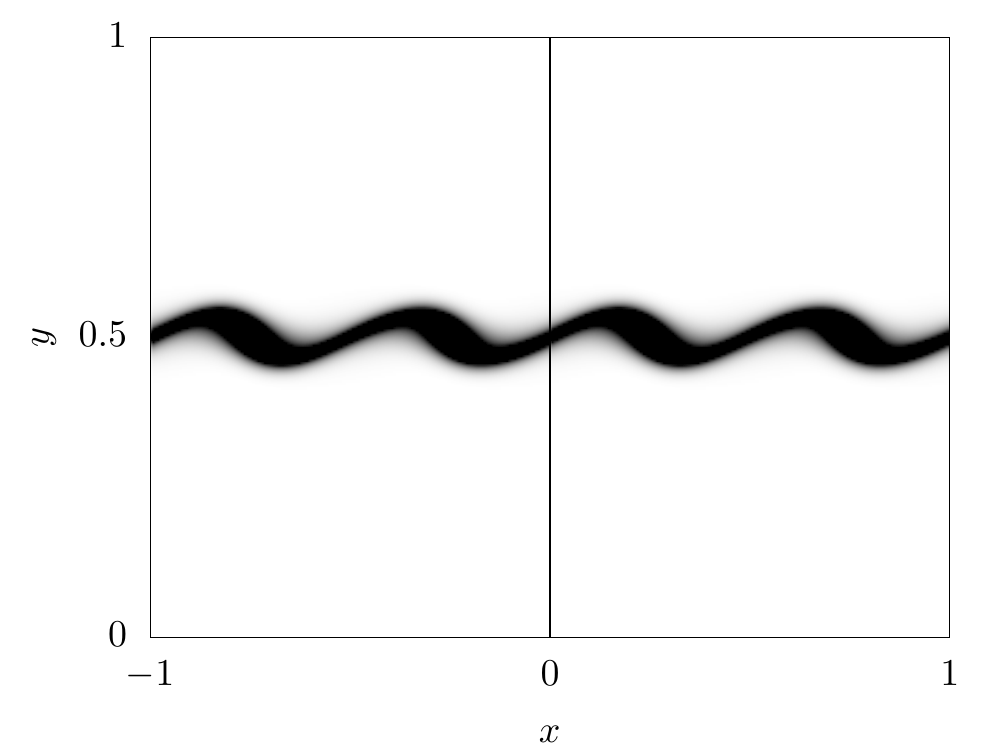}
\includegraphics[width=0.40\textwidth]{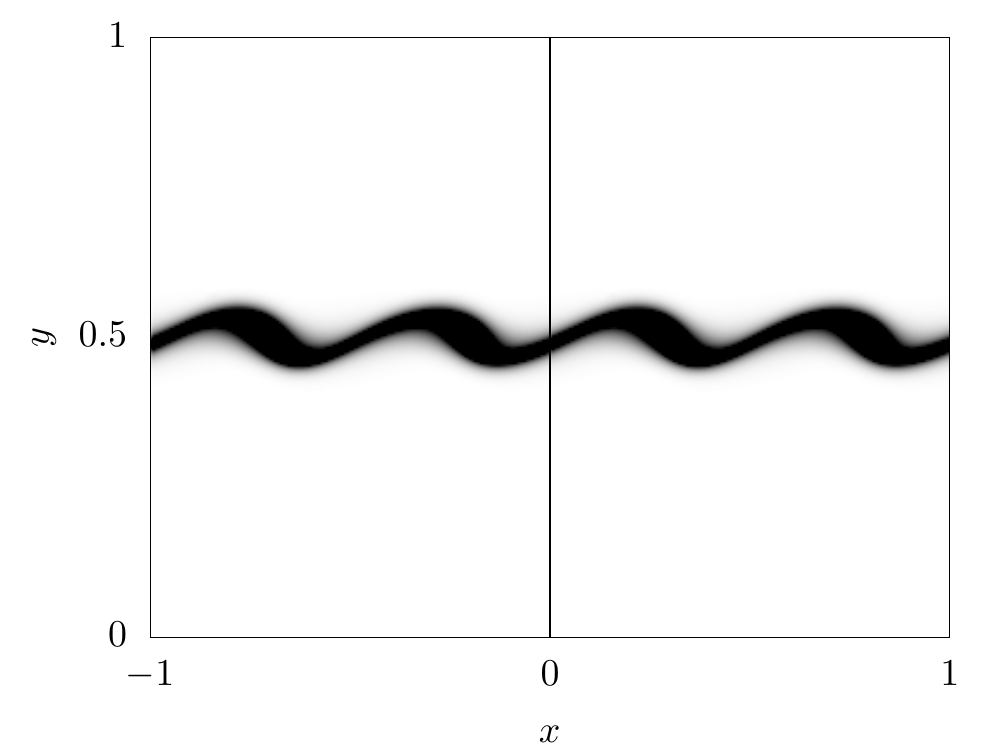}
\includegraphics[width=0.40\textwidth]{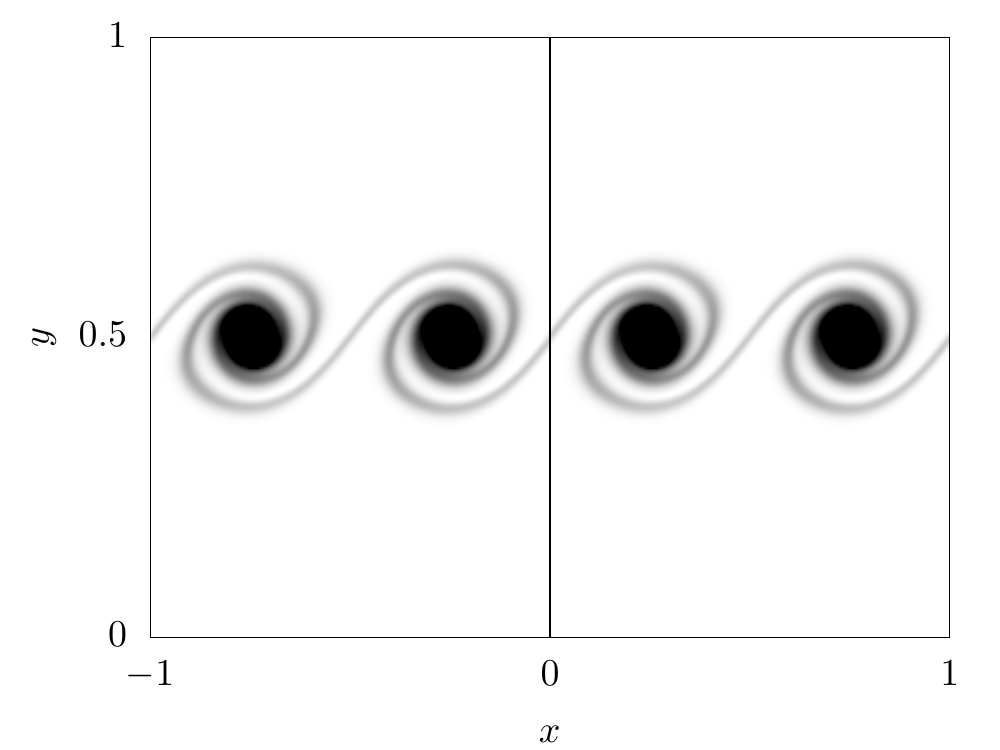}
\includegraphics[width=0.40\textwidth]{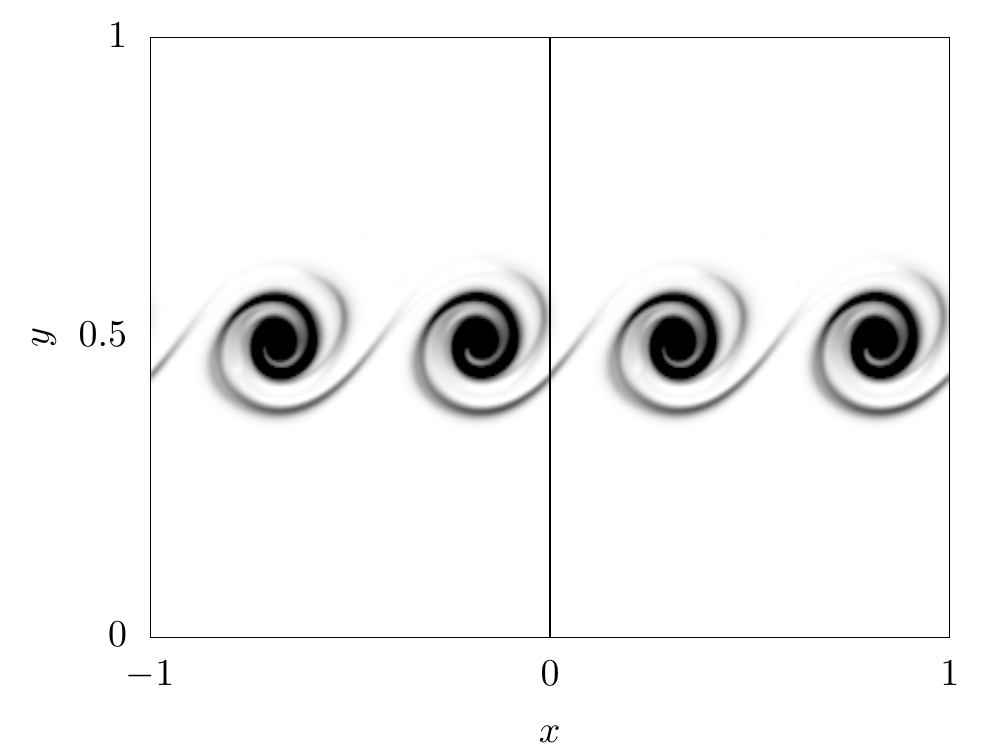}
\includegraphics[width=0.40\textwidth]{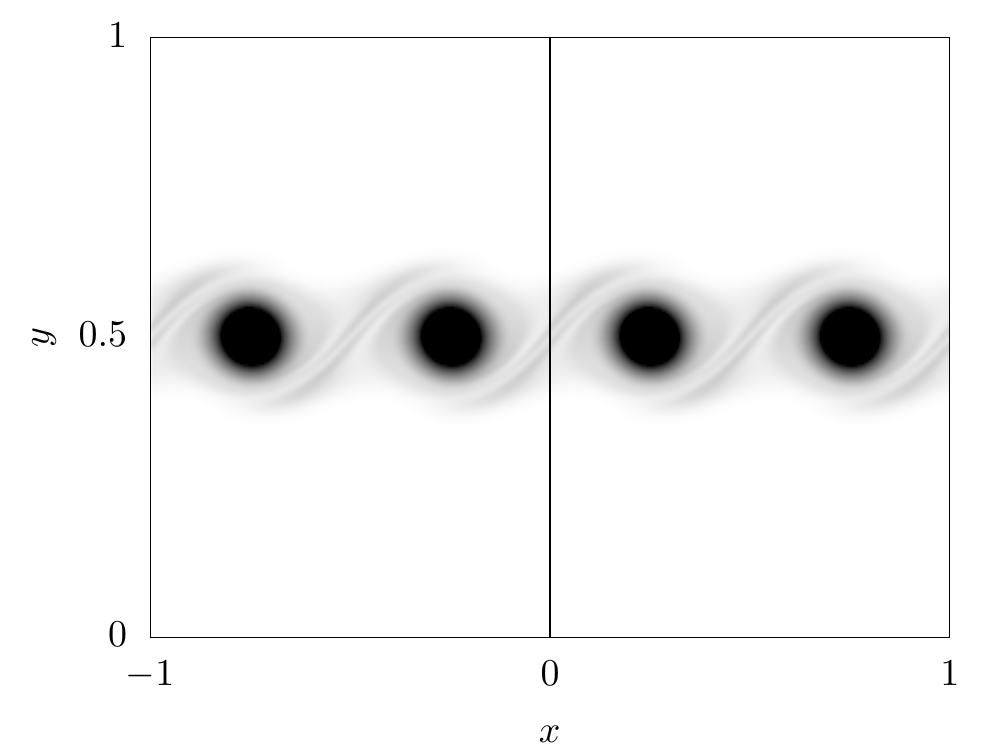}
\includegraphics[width=0.40\textwidth]{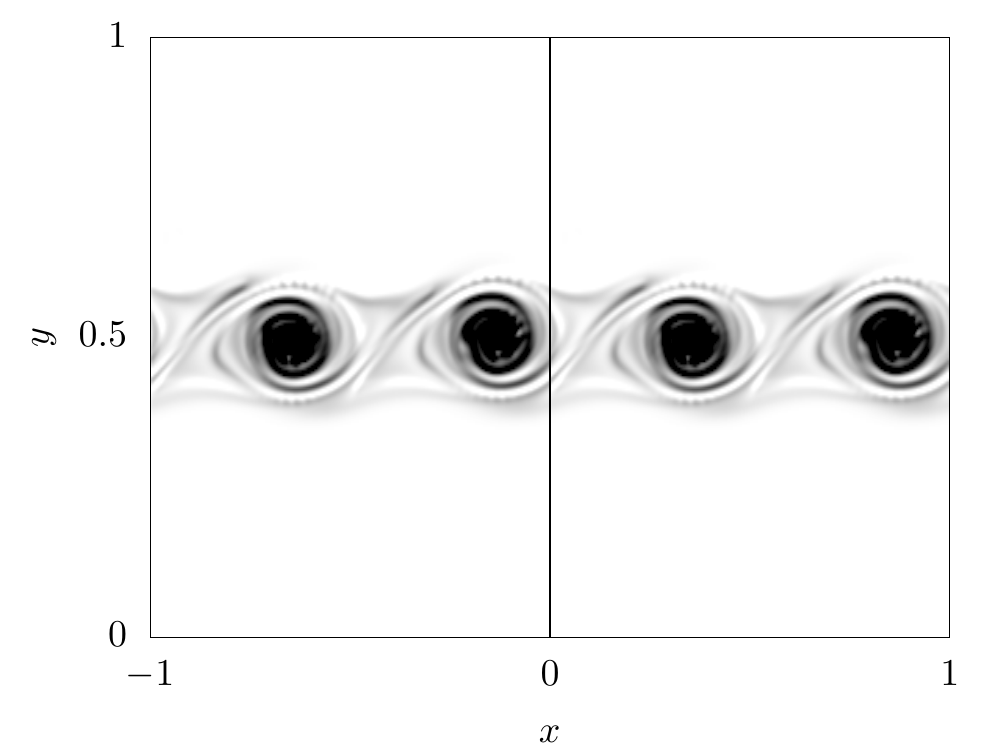}
\caption{Contour plots of the dimensionless vorticity field $\omega_z/\tilde{\omega}_c$ in the mixing layer for the isothermal case (left panels) and for the $(T_g/T_l)_i=3/4$ case (right panel).The contour plots refer to the dimensionless physical time $t/\tilde{t}=0.01$, $0.02$ and $0.03$, in order. The reference time scale is $\tilde{t}=\tilde{L}/U_c$ whereas the vorticity one is $\tilde{\omega}_c=U_c/\tilde{L}$.}
%\caption{Contour plots of the dimensionless vorticity field $\omega_z/\tilde{\omega}$ in the mixing layer for the isothermal case (left panels) and for the $(T_g/T_l)_i=3/4$ case (right panel).The contour plots refer to the dimensionless physical time $t/\tilde{t}=0.02$, $0.055$ and $0.075$, in order. The reference time scale is $\tilde{t}=\tilde{L}/U_c$ whereas the vorticity one is $\tilde{\omega}=U_c/\tilde{L}$. (CHECK!!! OMEGA DIMENSIONALE O NON-DIMENSIONALE?)}
\label{fig_c4_3}
\end{figure}
% FFFFFFFFFFFFFFFFFFFFFFFFFFFFFFFFFFFFFF
%
In the isothermal case, the thermodynamic pressure, the mean liquid and gas temperature and the mean gas density do not change over time as shown in Fig.~\ref{fig_c4_2}. Moreover with the prescribed boundary conditions and in absence of external forces, the mean kinetic energy in the incompressible case ( e.g. $\left(T_g/T_l\right)=1$) monotonically decreases due to the internal dissipation in the flow. On the other hand, as the temperature ratio is reduced below unity, the turbulent mixing enhances the thermal diffusion between the two fluids thus rapidly reducing the temperature gradients. As a result, the temperature tends to rapidly increase in the colder, compressible stream until a stationary condition is established. As the compressible phase heats up, the thermodynamic pressure, $p_0$, and the mean gas density increase. These effects modify the mean kinetic energy balance, that in the compressible cases contains not only the viscous dissipation but also a pressure work term proportional to the gas expansion. This last term modifies the variation of $E_k/E_{k,i}$ for all the compressible cases and is responsible for the initial increase in the mean kinetic energy observed for the case $\left(T_g/T_l\right)_i=0.75$ up to $t/\tilde{t}\approx 1$, $\tilde{t}=\tilde{L}/U_c$ being the reference time scale. However, once the temperature gradients become negligible and the thermodynamic pressure has reached a constant value, the compressible effects expire and the mean kinetic energy variation is mainly governed by the viscous dissipation. \par 
\scapin{Finally, Fig.~\ref{fig_c4_3} displays the contour plots of the instantaneous vorticity field at three different physical times for isothermal case and for the initial temperature ratio $\left(T_g/T_l\right)_i=3/4$. Despite we limit the analysis at the initial times in order to avoid the loss in resolution induced by the formation of smaller and smaller scales, we observe that, in general, the presence of a temperature gradient enhances the mixing and the growth-rate of the vorticity thickness with respect to the reference, isothermal case.}
\FloatBarrier
%
% ====================================================================================
%5
\section{Final remarks}
Multiphase, compressible flows are of great interest in a wide range of scientific fields and engineering problems. In this context, we propose a novel approach to the numerical simulation of multiphase, viscous flows where a compressible gas phase and an incompressible liquid mutually interact in the low-Mach number regime. The problem is addressed in the framework of a low-Mach number asymptotic expansion of the compressible formulation of the Navier-Stokes equations. In this limit, acoustics is neglected but large density variations of the gas phase can be accounted for as well as heat transfer between the phases and with the domain boundaries. A Volume of Fluid approach is used to deal with the presence of different phases in the flow as well as for interface tracking. The interface reconstruction is based on the MTHINC method~\cite{Ii2012interface} while the effect of the surface tension is accounted for using the continuum surface force (CSF) model~\cite{brackbill1992continuum}. The same set of equations is used for both the gas and the liquid phase, the zero-divergence condition being exactly imposed to the latter. To numerically solve this set of equations, we have developed a massive parallel solver, second order accurate both in time and space. The Poisson pressure equation is managed by a FFT-based solver that allows for a numerically efficient and very fast solution procedure. In addition, this choice is suited for code optimization and adaptation of incompressible GPU codes that benefits of FFT-based solvers (e.g. see~\cite{costa2020gpu}). The proposed iterative procedure shows to be more efficient in terms of number of iterations than the two approaches available in literature in the context of low-Mach number flows~\cite{motheau2016high,bartholomew2019new}. The solver has been build upon a code for incompressible flows which has undergone an extensive validation campaign~\cite{picano2015turbulent,rosti2017numerical,rosti2018rheology}. We provide a detailed and complete description of the theoretical approach together with information about the numerical technique and implementation details. In addition, we apply the described numerical approach to the simulation of five different flow configurations. The outcomes of two simulations reproducing the two-dimensional expansion and contraction of rectangular gaseous bands enclosed in an incompressible fluid and confined in a free-slip, periodic channel are provided. Next, we address the simulation of two-dimensional rising bubbles. First, we consider a single bubble and compare the results of our simulation with the reference data by Hysing et al.~\cite{hysing2009quantitative} using as benchmark quantities the bubble centroid and the bubble rising velocity. Second, we simulate the evolution of three bubbles of the same size but with different initial temperatures. Finally we discuss the outcome of a numerical simulation reproducing a plane, temporal mixing layer and show how the compressibility of the gas phase alters the development of the instability. \par 
As the proposed mathematical and numerical framework is independent of the capturing/tracking technique used to describe the interface topology, the proposed methodology can be directly extended to other existing numerical codes. We believe that the results presented here demonstrate that it is possible to accurately address the numerical simulation of multiphase, viscous flows in the low-Mach number regime, also when one of the phases can be treated as incompressible. Further extensions of the present methodology may concern the addition of more complex physical phenomena like phase change and complex interfacial thermodynamics, as absorption-desorption processes.
%
% ====================================================================================
%
\section*{Acknowledgements}
N.S., M.E.R. and L.B. acknowledge the support from the Swedish Research Council via the multidisciplinary research environment INTERFACE, Hybrid multiscale modelling of transport phenomena for energy efficient processes and the Grant No. 2014-5001. The computational resources were provided by SNIC (Swedish National Infrastructure for Computing).
\bibliography{mybibfile}
\end{document}